\documentclass[article,twocolumn,superscriptaddress]{revtex4-1}

\usepackage{amsmath}
\usepackage{amssymb}
\usepackage{amsthm}
\usepackage{graphicx}
\usepackage{url}
\usepackage{mathbbol}
\usepackage{mathtools}
\usepackage{xcolor}

\newcommand{\tr}[0]{\textrm{tr}}
\newcommand{\bra}[1]{\langle #1 |}
\newcommand{\ket}[1]{| #1 \rangle}
\newcommand{\braket}[2]{\langle #1 | #2 \rangle}

\begin{document}


\title{Quantum non-Gaussianity and secure quantum communication}

\author{Jaehak Lee}
\affiliation{Department of Physics, Texas A \& M University at Qatar, P.O. Box 23874, Doha, Qatar}
\author{Jiyong Park}
\affiliation{School of Basic Sciences, Hanbat National University, Daejeon 34158, Korea}
\author{Hyunchul Nha}
\email{hyunchul.nha@qatar.tamu.edu}
\affiliation{Department of Physics, Texas A \& M University at Qatar, P.O. Box 23874, Doha, Qatar}
\begin{abstract}

No-cloning theorem, a profound fundamental principle of quantum mechanics, also provides a crucial practical basis for secure quantum communication. The security of communication can be ultimately guaranteed if the output fidelity via communication channel is above the no-cloning bound (NCB). In quantum communications using continuous-variable (CV) systems, Gaussian states, more specifically, coherent states have been widely studied as inputs, but less is known for non-Gaussian states. We aim at exploring quantum communication covering CV states comprehensively with distinct sets of unknown states properly defined.
Our main results here are (i) to establish the NCB for a broad class of quantum non-Gaussian states including Fock states, their superpositions and Schrodinger-cat states and (ii) to examine the relation between NCB and quantum non-Gaussianity (QNG). We find that NCB typically decreases with QNG. 
Remarkably, this does not mean that quantum non-Gaussian states are less demanding for secure communication. By extending our study to mixed-state inputs, we demonstrate that QNG specifically in terms of Wigner negativity requires more resources to achieve output fidelity above NCB in CV teleportation. The more non-Gaussian, the harder to achieve secure communication, which can have crucial implications for CV quantum communications.
\end{abstract}

\maketitle

\section{\label{sec:introduction}Introduction}

No-cloning theorem is one of the fundamental quantum principles also providing a crucial practical basis for quantum communication---an eavesdropper cannot gain information without disturbing the quantum state carrying information.  Numerous works studied an approximate cloning scheme making clones with imperfect quality \cite{bib:DVCloning, bib:CVCloning1, bib:CVCloning2, bib:RMPCloning} and rigorously established as a security benchmark the no-cloning bound (NCB) above which the output fidelity of two clones cannot reach. \cite{bib:RMPCloning, bib:DVBound1, bib:DVBound2, bib:GaussianBound1, bib:GaussianBound2, bib:NonGaussianBound}. If a receiver obtains an output state with fidelity higher than NCB, he can be assured of no better copy existing elsewhere and extract more information than Eve---an ultimate security of communication. 
Such a connection was specifically made between the optimal cloning and the security of quantum cryptographic protocols \cite{bib:RMPCloning, bib:DVBound2, bib:DVCrypt, bib:CVCrypt, bib:CVQKD}. 

The NCB varies with the set of input states and it is of crucial importance to identify it for different input states to address communication security relevant to various protocols. 
In quantum communication using continuous variables (CVs),  coherent states are readily available information carriers and have been mostly employed as input states to many protocols, e.g. quantum cryptography \cite{bib:CVQKD} and quantum teleportation \cite{bib:BKteleportation}. Furthermore, it was recently proved that coherent states are the optimal input states achieving the ultimate classical capacity of bosonic Gaussian channels \cite{bib:MinimumOutputEntropy1, bib:MinimumOutputEntropy2}. The NCB of coherent states was well studied with some extension to other Gaussian (squeezed) states \cite{bib:GaussianBound1, bib:GaussianBound2,bib:GQI}. On the other hand, quantum non-Gaussian states have recently drawn much attention as an essential ingredient for quantum information processing due to the limited capability of Gaussian states and operations in some crucial tasks, e.g. entanglement distillation \cite{bib:distillation}, quantum computation \cite{bib:computation}, and error correction \cite{bib:errorcorrection}.  However, little is known about non-Gaussian states particularly their NCBs and the critical role played by their quantum non-Gaussianity (QNG) in quantum communications. QNG here refers to the characteristic of non-Gaussian states that cannot be represented as a mixture of Gaussian states, with its measure rigorously quantifying distinction between those two sets of states \cite{bib:QNGmeasure}. For instance, some studies investigated several non-Gaussian input states for CV teleportation \cite{bib:NGinput1, bib:NGinput2, bib:NGinput3}, which however lacks the analysis of performance in view of ultimate security with no access to NCB benchmark. We need to extend our approach to deal with a broad class of quantum non-Gaussian states and to rigorously identify NCB and its relation to QNG in quantum communications. By doing so, we may address some important issues e.g. comparing performance and security conditions between Gaussian and non-Gaussian states in CV quantum information protocols  \cite{bib:robustness1, bib:robustness2, bib:robustness3, bib:robustness4, bib:TeleportationNGResource1, bib:TeleportationNGResource2, bib:PhysRevLett.108.030503, bib:PhysRevA.95.052343}.


Our objective is two-fold. First, we intend to establish NCB for quantum non-Gaussian states broadly to gain insight into non-Gaussian regime. 
Second, we intend to identify the role played by QNG for CV quantum communications. There have recently been a growing interest in studying QNG as a resource under the framework of resource theories \cite{bib:WignerNegativity1, bib:WignerNegativity2, bib:QNGmeasure}, with its operational significance largely unexplored \cite{bib:operational}. We here find a strong correlation between QNG and NCB---NCB overall decreases with QNG. Importantly, this does not mean that achieving ultimate security becomes easier with quantum non-Gaussian states.  We show that non-Gaussian input states require more resources to achieve secure teleportation  \cite{bib:BKteleportation, bib:SecureTeleportation1, bib:SecureTeleportation2} even though NCB is smaller. This implies that the security of CV teleportation can be most readily attained when employing Gaussian input states. 


\section{Results}
To address CV states comprehensively and systematically, we first define a set of states as ${\cal S}_{|\psi\rangle}\equiv\{{\hat D}(\alpha)|\psi\rangle~{\rm for}~{\forall} \alpha\in C\}$, where each set $\cal S_{|\psi\rangle}$ consists of a pure state $|\psi\rangle$ arbitrarily displaced in phase space ${\hat D}(\alpha)\equiv e^{\alpha a^\dag-\alpha^*a}$ (Fig. 1 (a)). The set of coherent states is a special case with $|\psi\rangle=|0\rangle$ (vacuum). In generalizing CV states this way, we may examine quantum non-Gaussian states comprehensively varying the state $|\psi\rangle$ for each set. Furthermore, displacement operation provides an important protocol of information encoding for CV communications. Quantum teleportation of coherent states corresponds to transmitting information on the unknown displacement in phase space for a given seed state $|0\rangle$. In addition, the ultimate classical capacity under Gaussian channels uses encoding based on displacement as an optimal scheme \cite{bib:MinimumOutputEntropy1, bib:MinimumOutputEntropy2}. A full displacement in phase space can represent an increasingly large amount of information.
	\begin{figure}[t]
	\centering \includegraphics[clip=true, width=0.9\columnwidth]{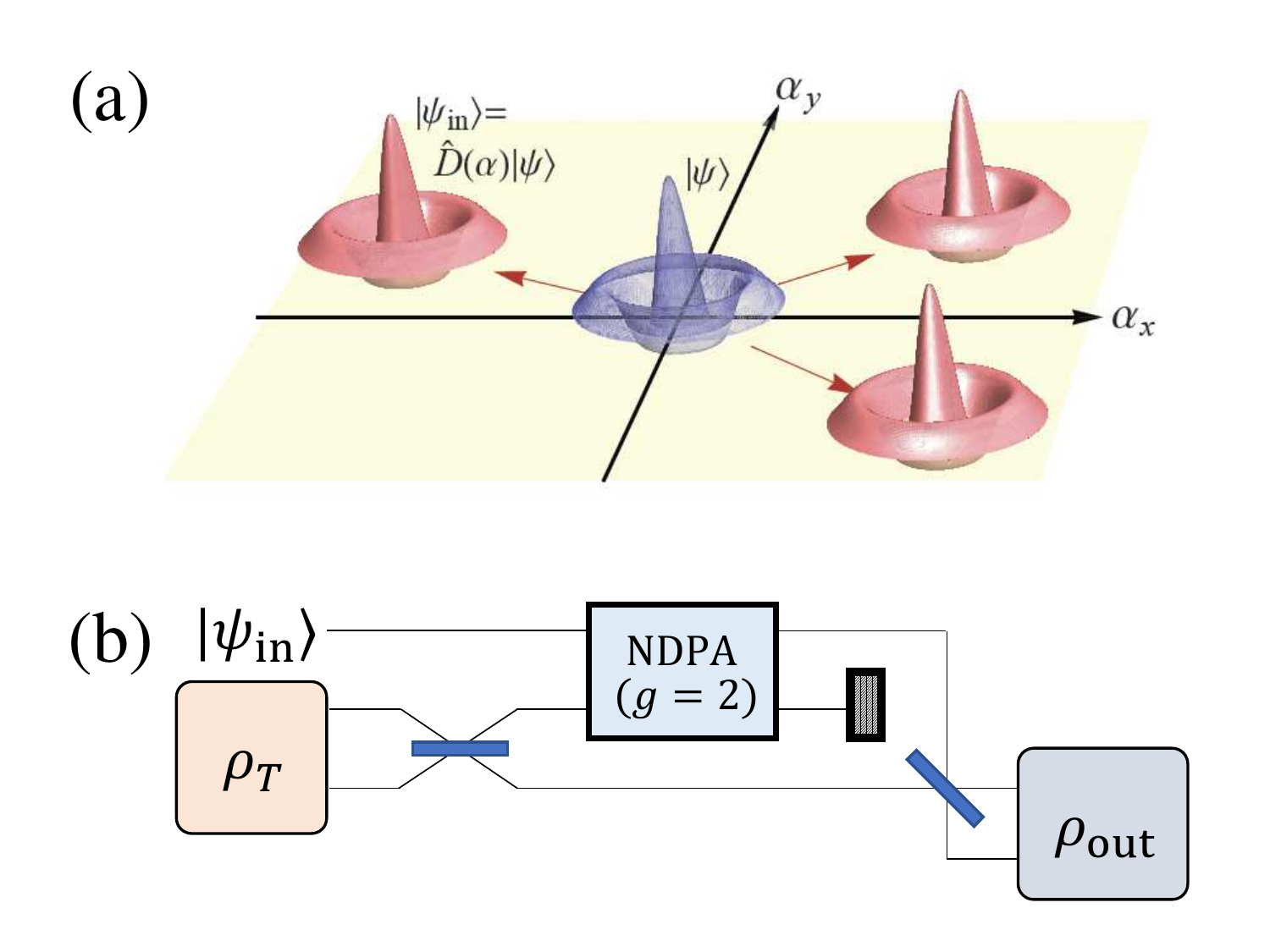}
	\caption{\label{fig:implementation} (a) A quantum state $|\psi\rangle$ is displaced by an unknown amplitude in phase space, which defines a set of states ${\cal S}_{|\psi\rangle}\equiv\{{\hat D}(\alpha)|\psi\rangle~{\rm for}~{\forall} \alpha\in C\}$. An example of a non-Gaussian state $|\psi\rangle=|2\rangle$ is shown. (b) Optimal $ 1 \to 2 $ cloner can be implemented by a non-degenerate parametric amplifier (NDPA) and 50/50 beamsplitters with an ancilla state $ \rho_\textrm{T} $ optimized.}
	\end{figure}

\textbf{Covariant cloning machine}--- We first investigate the cloning of Fock states (prominently quantum non-Gaussian) under displacement, i.e. the sets ${\cal S}_{|\psi\rangle=\ket{n}}$, which will be extended to other non-Gaussian states like superposition of Fock states and Schrodinger cat-states later. 
Our goal is to maximize the output fidelity averaged over all input displacements $ \alpha $ for a given set ${\cal S}_{|n\rangle}$. For simplicity we assume that the distribution of unknown $ \alpha $ is uniform in the whole phase space. For an 1-to-$M$ cloning map $ T $, the fidelity of $i$th clone is expressed as
	\begin{equation} \label{eq:singlefidelity}
	F^{(i)} = {\rm Tr} \left[ T(\rho_\textrm{in}) \rho_\textrm{in}^{(i)} \right] ,
	\end{equation}
where $ \rho^{(i)} \equiv \mathbb{1} \otimes \cdots \otimes \mathbb{1} \otimes \rho \otimes \mathbb{1} \otimes \cdots \otimes \mathbb{1}  $ has its component $ \rho $ on $i$th system whereas others are given by identity operator $ \mathbb{1} $.
Let us consider a symmetric cloner with identical fidelity for each $ i $. The optimal cloner can then be given by a covariant cloner \cite{bib:NonGaussianBound, bib:Kruger} with the covariance property $ T (\rho) = T_\beta (\rho) $ regardless of $\beta$  (Methods), where 
	\begin{equation} \label{eq:covariantcloner}
	T_\beta (\rho) = \hat{D}^\dagger{}^{\otimes M} (\beta) ~ T \left[ \hat{D} (\beta) \rho \hat{D}^\dagger (\beta) \right] \hat{D}^{\otimes M} (\beta)
	\end{equation}
is a shifted cloner. If an input state is displaced by $ \beta $, each output state is also displaced by the same $ \beta $. 
With this covariance,  input and output states must satisfy the relation in terms of the characteristic function $ \chi (\xi) = \tr [ \rho \hat{D} (\xi) ] $ \cite{bib:Barnett} as \cite{bib:NonGaussianBound, bib:Kruger}
	\begin{equation} \label{eq:covariantinout}
	\chi_\textrm{out} ( \boldsymbol{\xi} ) = \chi_\textrm{T} ( \Omega \boldsymbol{\xi} ) \chi_\textrm{in} ( {\textstyle \sum_i} \xi_i ).
	\end{equation}
Here $ \Omega $ is a linear transformation acting as $ \xi_{i,p} \to \xi_{i,x} $ and $ \xi_{i,x} \to \sum_{j \neq i} \xi_{j,p} $ with $ \boldsymbol{\xi} = ( \xi_{1,x}, \xi_{1,p}, \cdots, \xi_{M,x}, \xi_{M,p} ) $ and $ \chi_\textrm{T} ( \boldsymbol{\xi} ) $ is the characteristic function of a certain $M$-mode state $ \rho_\textrm{T} $. That is, our problem is reduced to finding an optimal state $ \rho_\textrm{T} $ giving a maximal fidelity through Eq. (1). This $ \rho_\textrm{T} $ can be utilized as a resource state to construct the optimal cloner \cite{bib:NonGaussianBound} and the equivalent telecloning protocol \cite{bib:PhysRevA.94.062318}. For instance, the optimal $ 1 \to 2 $ cloner can be realized using a scheme in Fig. 1 (b).

Using Eq. (\ref{eq:covariantinout}), the fidelity in (\ref{eq:singlefidelity}) for each $n$ is given, with details in Supplementary Section I, by
	\begin{equation} \label{eq:fj}
	F_n^{(i)} \equiv \left\langle \hat{f}_n^{(i)} \right\rangle_{\rho_\textrm{T}} = \left\langle [ L_n ( \hat{O}_i ) ]^2 e^{-\hat{O}_i} \right\rangle_{\rho_\textrm{T}} ,
	\end{equation}
where $ \hat{O}_i \equiv \frac{1}{2} \left[ \hat{x}_i^2 + ( \sum_{j \neq i} \hat{p}_j )^2 \right] $ with $ \hat{x}_i $ and $ \hat{p}_i $ being position and momentum operators of $i$th mode, respectively, and $L_n$ the Laguerre polynomial of order $n$. 
NCB of our interest is given by considering the case of two clones ($M=2$) with the average fidelity $\frac{1}{2}(F_n^{(1)}+F_n^{(2)})$ optimized over a resource state $\rho_\textrm{T}$.
That it, NCB corresponds to the maximum eigenvalue of the operator $ \frac{1}{2}(\hat{f}_n^{(1)}+\hat{f}_n^{(2)})$ [See Supplementary Section I]. For instance, the NCB$\sim 0.6826 $ for coherent-state inputs ($n=0$) is given by the maximum eigenvalue of  $ \frac{1}{2} ( e^{-\frac{1}{2}(\hat{x}_1^2+\hat{p}_2^2)} + e^{-\frac{1}{2}(\hat{x}_2^2+\hat{p}_1^2)} ) $ \cite{bib:NonGaussianBound, bib:Kruger}.

{\it Invariance under Gaussian unitaries:}~Remarkably, NCB is invariant under Gaussian unitary operations. Namely, NCBs for the two sets ${\cal S}_{|\psi\rangle}$ and ${\cal S}_{{\hat U}_\textrm{G}|\psi\rangle}$ are identical, where ${\hat U}_\textrm{G}$ is an arbitrary Gaussian unitary (Methods). Thus, all pure Gaussian states attain the same NCB, which was mentioned for the Gaussian cloners using only Gaussian resource states \cite{bib:GaussianBound1, bib:GaussianBound2}, but the same is true with general covaiant cloners using non-Gaussian resources for any input states. For instance, the Fock states and the squeezed Fock states have the same NCBs, which make our result encompass non-Gaussian states more broadly.

\textbf{Gaussian No-cloning Bound}---Before we deal with the ultimate cloning limit, 
it is worth investigating NCB under the constraint of Gaussian operations because Gaussian operations are highly feasible in laboratory \cite{bib:GQI}. $\rho_\textrm{T}$ in Eq. (2) is then a Gaussian state with $ \chi_\textrm{T} ( \Omega \boldsymbol{\xi} ) = \exp \left( -\frac{1}{2} \boldsymbol{\xi}^\mathrm{T} \gamma_t \boldsymbol{\xi} \right) $ and $ \gamma_t = a \mathbb{1}_M \otimes \mathbb{1}_2 + b ( \mathbb{E}_M - \mathbb{1}_M ) \otimes \mathbb{1}_2 $ for a symmetric covariant cloner [See Supplementary Section III]. ($ \mathbb{1}_d $: $ d \times d $ identity matrix, $ \mathbb{E}_d $: $ d \times d $ matrix with $ (\mathbb{E}_d )_{i,j} = 1 $ for all $ i, j $). The positivity of the cloning transformation is fulfilled if and only if $ a-b \ge 1 $ and $ a + (M-1)b \ge M-1 $. We find that the maximum fidelity is achieved, interestingly regardless of Fock states, at $ a = \frac{2M-2}{M} $ and $ b = \frac{M-2}{M} $, with details in Supplementary Section III. For $ M = 2 $, we obtain the Gaussian NCB $ F_0^\textrm{(G)nc} = \frac{2}{3} \simeq 0.6667 $, $ F_1^\textrm{(G)nc} = \frac{10}{27} \simeq 0.3704 $, $ F_2^\textrm{(G)nc} = \frac{22}{81} \simeq 0.2716 $, and $ F_3^\textrm{(G)nc} = \frac{490}{2187} \simeq 0.2241 $, which rapidly decreases with $n $.

\textbf{Ultimate No-cloning Bound}---
Although Gaussian NCB is a useful benchmark, the ultimate NCB must be obtained by optimizing over all possible physical operations. The problem is to find the maximum value of $ \frac{1}{2} \big\langle \hat{f}_n^{(1)} + \hat{f}_n^{(2)} \big\rangle_{\rho_\textrm{T}} $ over all quantum states $ \rho_\textrm{T} $. We examine the largest eigenvalue of the corresponding operator in the truncated Fock-state basis $ \ket{n_1, n_2} $ with $ 0 \le n_1, n_2 \le N_\textrm{trunc} $. The density matrix elements of (\ref{eq:fj}) can be evaluated using position(momentum) representation of Fock states [See Supplementary Section I]. We numerically obtain the ultimate NCB decreasing with $n$ as $ F_0^\textrm{nc} \simeq 0.6826 $, $ F_1^\textrm{nc} \simeq 0.5449 $, $ F_2^\textrm{nc} \simeq 0.5145 $, and $ F_3^\textrm{nc} \simeq 0.5033 $ with $ N_\textrm{trunc} = 300 $. We have checked that our results are stable with negligible change over varied $ N_\textrm{trunc} $ [See Supplementary Figure 1]. Our method also gives the same known NCB for the coherent-state inputs in \cite{bib:NonGaussianBound}.

For non-Gaussian states, there is a substantial difference between the ultimate NCB and the Gaussian NCB, whereas the difference is quite small for Gaussian states.  For the coherent-state input, while each component operator $ \hat{f}_0^{(i)} $ to determine fidelity is Gaussian, the sum of those two operators $ \frac{1}{2}(\hat{f}_0^{(1)}+\hat{f}_0^{(2)})= \frac{1}{2} ( e^{-\frac{1}{2}(\hat{x}_1^2+\hat{p}_2^2)} + e^{-\frac{1}{2}(\hat{x}_2^2+\hat{p}_1^2)} ) $ is non-Gaussian, which makes the non-Gaussian cloning an optimal cloner. For non-Gaussian input states, on the other hand, each component operator $ \hat{f}_n^{(i)} $ itself is already non-Gaussian, not to mention their sum. While Gaussian states also require non-Gaussian resources for optimal cloning \cite{bib:NonGaussianBound}, our result clearly shows that non-Gaussian resources are more essential to optimally clone non-Gaussian inputs.

\textbf{Secure Teleportation and QNG}---
We now demonstrate that the smaller NCB does not merely mean `easy to achieve' by examining an important communication protocol---CV quantum teleportation \cite{bib:BKteleportation}. 
The output state of CV teleportation can be described by $ \chi_\textrm{out} ( \xi ) = \chi_\textrm{in} ( \xi ) \chi_\textrm{AB} ( \xi^\ast, \xi ) $ where $ \chi_\textrm{AB} ( \xi_1, \xi_2 ) $ is the characteristic function of the two-mode resource state \cite{bib:PhysRevLett.108.030503, bib:PhysRevA.74.042306}. Let us consider a typical resource, i.e. two-mode squeezed state (TMSV) to illustrate a general trend. The teleportation fidelity is then given by $ F = \frac{1}{\pi} \int d^2\xi \chi_\textrm{in} ( -\xi ) \chi_\textrm{in} ( \xi ) e^{-e^{-2r} |\xi|^2 } $ with $ r $ squeezing parameter. Both of the mean photon number and the entanglement of TMSV increases with $ r $, which can be used as a measure of required resource. We investigate the critical value of $ r $ to achieve secure teleportation above the NCB for each case.

(i) {\it Fock-state inputs}---Interestingly, the output fidelity above the Gaussian NCB can be achieved with the same squeezing $ r = \tanh^{-1} \frac{1}{3} $ regardless of Fock states $|n\rangle$. In this case, the correlated quadratures of TMSV become $ \left\langle \Delta^2 ( \frac{\hat{x}_\textrm{A} - \hat{x}_\textrm{B}}{\sqrt{2}} ) \right\rangle = \left\langle \Delta^2 ( \frac{\hat{p}_\textrm{A} + \hat{p}_\textrm{B}}{\sqrt{2}} ) \right\rangle = \frac{1}{2} V_0 $, where $ V_0 $ is the vacuum fluctuation. This result reflects the feature of Gaussian cloners \cite{bib:GaussianBound1}, where the optimal cloning is achieved when the added noise is the half of vacuum fluctuation.
To achieve the fidelity corresponding to the ultimate NCB, resource requirement $r$ is of course higher than that for the Gaussian NCB. In Fig. 2, we see that higher squeezing $r$ is needed for a larger $ n $, even though the NCB itself is smaller.
	\begin{figure}[t]
	\centering \includegraphics[clip=true, width=0.8\columnwidth]{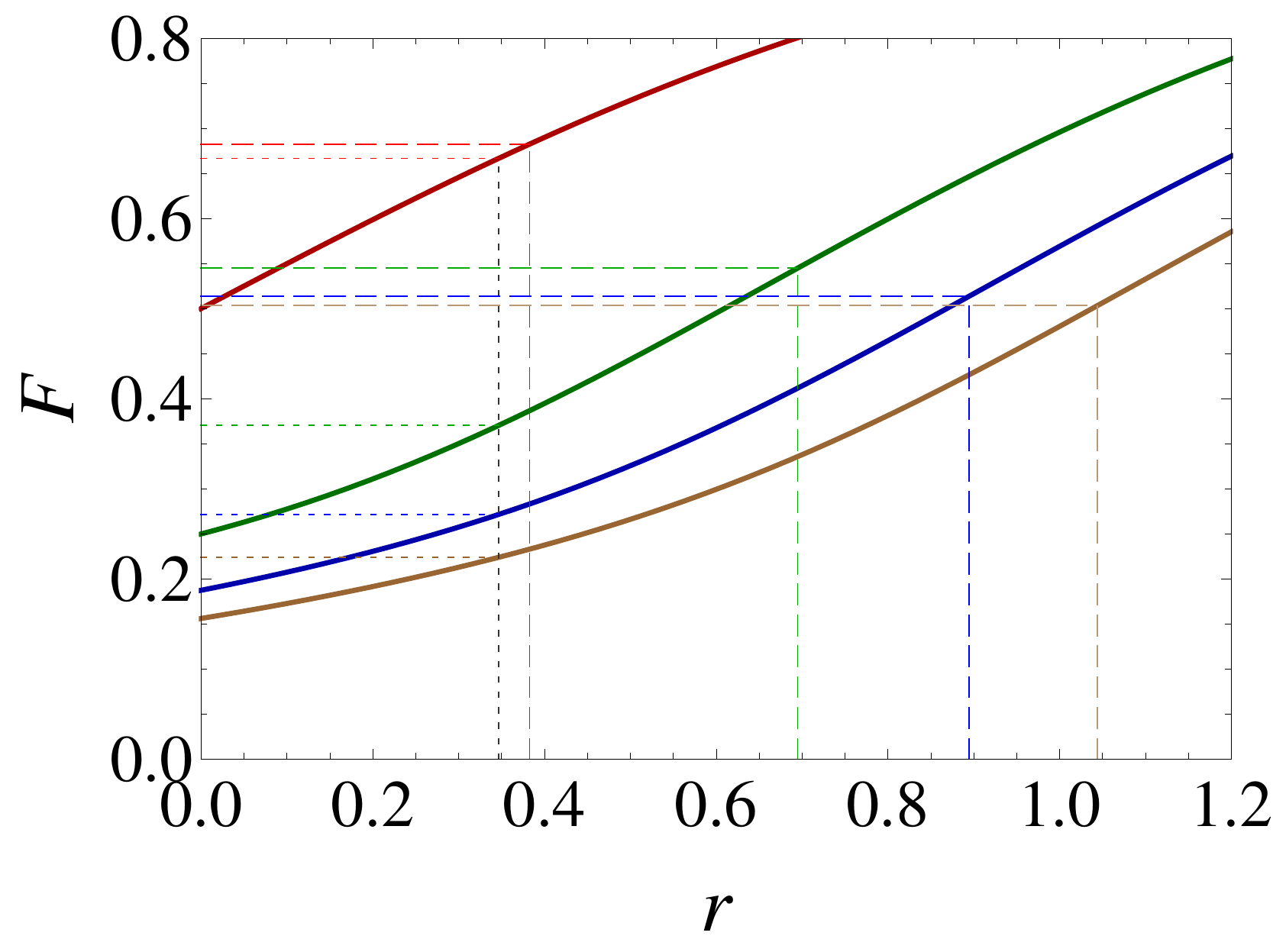}
	\caption{\label{fig:Fockfidelity} Teleportation fidelity using TMSV as a resource [Supplementary Equation (32)] and squeezing requirement to achieve NCB. Solid curves represent the fidelity for $ n = $ 0(red), 1(green), 2(blue), 3(brown) from top to bottom. Horizontal dashed (dotted) lines represent the ultimate (Gaussian) NCB and the vertical dashed (dotted) lines the required squeezing to achieve each NCB.}
	\end{figure}

One may wonder if our finding varies with the type of resource state used for teleportation, as we have used a Gaussian resource, TMSV, even for the teleportation of non-Gaussian states. It was shown that the teleportation fidelity can be improved by employing non-Gaussian resources under energy constraint for coherent-state inputs \cite{bib:PhysRevA.95.052343}. In Supplementary Section VI, we perform a similar analysis for non-Gaussian input states to draw the same conclusion, i.e. teleportation of non-Gaussian input states requires more resources to overcome the NCB even when optimal non-Gaussian entangled resources are employed.

As a remark, the output state via teleportation with TMSV is given by $ \chi_\textrm{out} ( \xi ) = \chi_\textrm{in} ( \xi ) e^{-e^{-2r} |\xi|^2 }$, which is equivalent to the output under Gaussian noise-added channels. Thus our result also represents the robustness of security under Gaussian channels, which is strongest when Gaussian input states are employed. 

It is also an interesting issue how Alice and Bob can actually confirm security in teleportation as the output fidelity above NCB should be verified with respect to {\it unknown} input states. In a typical teleportation experiment \cite{bib:CVTreview}, there exists a third party who provides an input state to Alice and who also checks the output fidelity afterwards to assess the performance quality. On the other hand, if Alice and Bob themselves want to do the security analysis, they may execute a trial teleportation for a subset of sample states. That is, Alice herself prepares some states (selected from the set of unknown states they are supposed to teleport) and performs the teleportation protocol with Bob who can check the fidelity with the information Alice gives. This procedure is somewhat similar to the security test in quantum key distribution \cite{bib:DVCrypt, bib:CVCrypt, bib:CVQKD} using sample data out of raw correlated data established between Alice and Bob, which enables them to analyze the properties of channel they use and bound the information Eve possesses.

 (ii) {\it general pure-state inputs}---We now want to discuss the relation between QNG and NCB for secure quantum communication more rigorously. A faithful measure of quantum non-Gaussianity was introduced in \cite{bib:QNGmeasure} using a convex-roof extension of non-Gaussianity defined by the relative entropy $ \delta (\rho) \equiv S ( \rho | \tau ) = S (\tau) - S (\rho) $  \cite{bib:NGmeasure1, bib:NGmeasure2} for a given state $ \rho $ with respect to its reference Gaussian state $ \tau $ having the same first and second moments. ($ S$: von Neumann entropy). For a pure state, QNG is simply given by the entropy of the reference Gaussian state $S (\tau)$. We here consider two general non-Gaussian states, a superposition of Fock states and cat states with results in Fig. 3 (a) showing NCB versus QNG .  The procedures to obtain NCB for general input states are described in Supplementary Section I. For superposition states $ c_0\ket{0} + c_1\ket{1} $, $c_0\ket{0} + c_2\ket{2} $ and $ c_1\ket{1} + c_2\ket{2} $, we have calculated each quantity by changing coefficients gradually. For a comprehensive analysis, we have also generated the superposition of $ \ket{0} $, $ \ket{1} $, and $ \ket{2} $ randomly $ 9 \times 10^3 $ times (grey dots). For even or odd cat state, $ \mathcal{N}_\alpha^\pm ( \ket{\alpha} \pm \ket{-\alpha} ) $ with $ \mathcal{N}_\alpha^\pm = \left( 2 \pm 2 e^{-2 \alpha^2} \right)^{-1/2} $ normalization factor, we obtain NCB and QNG by changing $ \alpha $ from 0 to 2.
	\begin{figure}[t]
	\centering \includegraphics[clip=true, width=0.9\columnwidth]{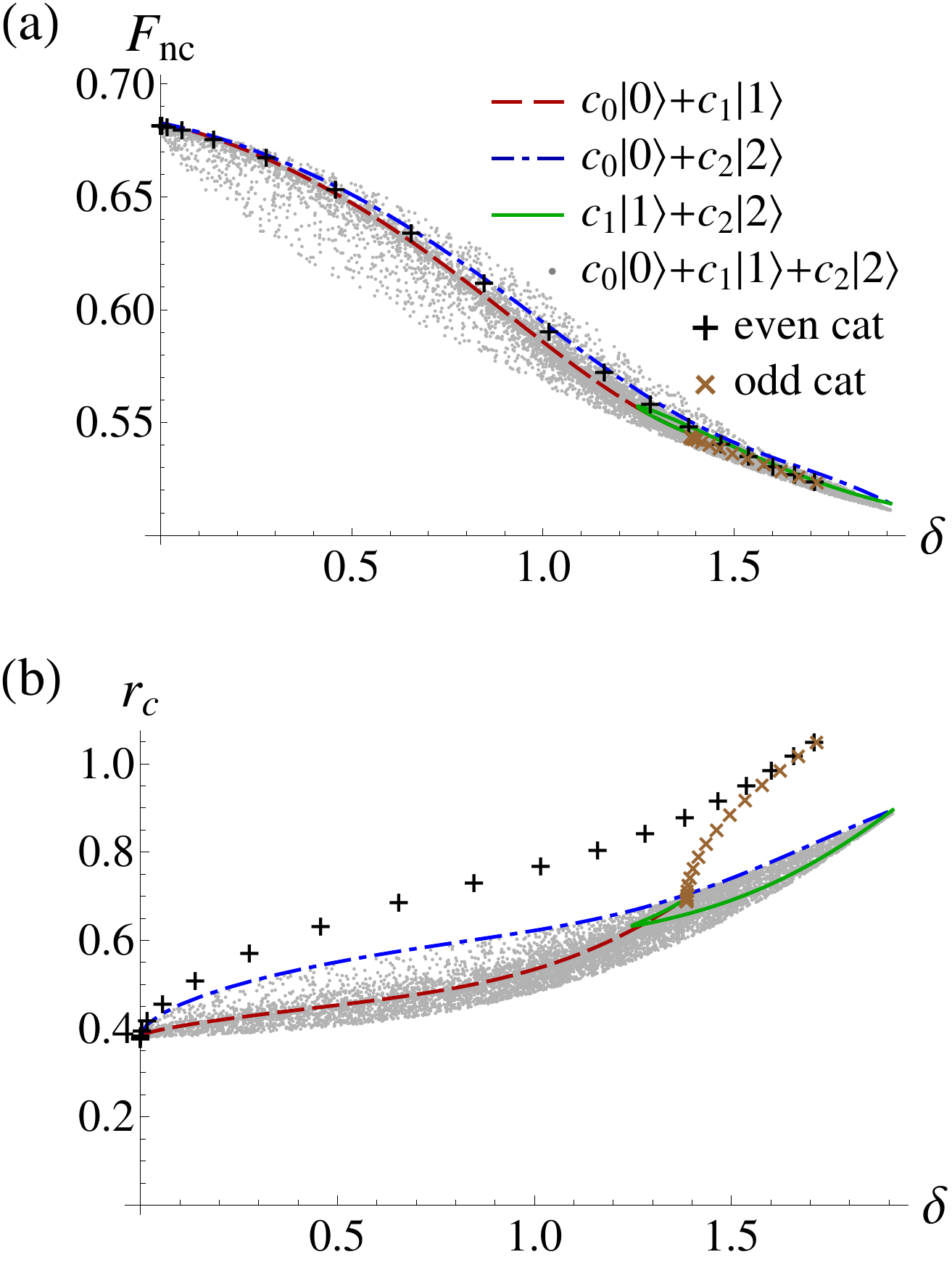}
	\caption{\label{fig:plotQNG} (a) QNG in terms of $\delta$ (main text) versus NCB for quantum non-Gaussian states. (b) QNG versus critical squeezing to achieve secure quantum teleportation.}
	\end{figure}

For a fixed QNG, different input states show very close value of NCB and we find an overall trend---NCB decreases with QNG. 
We also note that the NCBs for all non-Gaussian states are below that of Gaussian states, 0.6826. 

We also plot the critical squeezing $r_\textrm{c}$ required to achieve NCB in Fig. 3 (b).
$r_\textrm{c}$ varies with a class of input states at a fixed QNG, however it shows a monotonic behavior within the same class. The difference might be further reduced when one employs resource states optimized for each input state. Our result shows that it becomes more demanding to achieve secure teleportation with a larger QNG of input states. We recall that the NCB is invariant under Gaussian unitaries, which means our result covers broader class of non-Gaussian states including squeezed non-Gaussian states. QNG $ \delta $ is also invariant under Gaussian unitaries \cite{bib:QNGmeasure}, which supports our conclusion in relation to QNG and NCB more broadly. 

(iii) {\it mixed-state inputs}--- So far we have considered pure-state inputs as an ideal information carrier. We intend to further extend this approach to mixed-state inputs to more rigorously identify which source of QNG is a critical factor affecting secure communication. 
For mixed-state inputs, we adopt a trace method to quantify the overlap of two states in Eq. (1)  instead of the usual fidelity measure. 
The overlap ${\rm Tr} \{\rho\tau\}$ may be operationally interpreted as the expectation value of the output state $\tau$ over the {\it observable} $\rho$ \cite{bib:Owari}. In another perspective, it represents the overlap of the Wigner functions in phase space, i.e., 
${\rm Tr} \{\rho\tau\}=\pi\int d^2\alpha W_\rho(\alpha)W_\tau(\alpha)$. This quasi-probability overlap can also be readily measured experimentally by mixing two states $\rho$ and $\tau$ at a 50:50 beam splitter and then observing the number parity of one output state. That is, $W_\rho(\alpha_1)W_\tau(\alpha_2)\xrightarrow{\text{BS}} W_{12}'(\alpha_1,\alpha_2)=W_\rho(\frac{\alpha_1+\alpha_2}{\sqrt{2}})W_\tau(\frac{\alpha_1-\alpha_2}{\sqrt{2}}) $, from which the parity of mode 2 is given by $\frac{\pi}{2}W_2'(\alpha_2=0)=\frac{\pi}{2}\int d^2\alpha_1 W_{12}'(\alpha_1,0)=\pi\int d^2\alpha W_\rho(\alpha)W_\tau(\alpha)$. This method was used, e.g., to detect entanglement by measuring local and global purities ${\rm Tr}\{\rho^2\}$ after preparing two identical states \cite{bib:Nature}.
This overlap measure applied to mixed states may involve an undesirable feature. For instance, given $\rho=0.75|0\rangle\langle0|+0.25|1\rangle\langle1|$, a different state $\tau=0.9|0\rangle\langle0|+0.1|1\rangle\langle1|$ may yield a higher overlap than $\rho$ itself, i.e. ${\rm Tr} \{\rho\tau\}>{\rm Tr} \{\rho^2\}$, an issue related to mixedness of each state. Nevertheless, with more details in Supplementary Section VII,  our overlap measure is relevant to capture the essence of quantum cloning and related problems.

Importantly, when both of the no-cloning bound and the performance quality of teleportation are assessed and compared using the same measure (fidelity or state-overlap), our analysis is fair: By its construction, the NCB derived in terms of trace in Eq. (1) (state-overlap in phase space) indicates that no pair of two cloned copies can possess the value above the bound in terms of the same measure. Therefore, for a given mixed input-state $ \rho_{\rm in} $, if the condition $ {\rm Tr} \{ \rho_{\rm tele} \rho_{\rm in} \} > {\rm Max}\frac{1}{2}\left[ {\rm Tr}\{\rho_{\rm clone1}\rho_{\rm in} \}+{\rm Tr}\{\rho_{\rm clone2}\rho_{\rm in}\}\right] $, where the latter maximum is our NCB benchmark, is verified importantly in ensemble sense averaged over unknown amplitudes $\alpha$ in the entire phase space,  we can say there does not exist a copy possessing better state-overlap than the teleported state with the original state.

To begin with, we need to make distinction between QNG and simple non-Gaussianity. Even a finite mixture of Gaussian states can be a non-Gaussian state, which however is not regarded as genuinely quantum non-Gaussian \cite{bib:GenuineNG1, bib:GenuineNG2, bib:GenuineNG3, bib:GenuineNG4, bib:GenuineNG5, bib:GenuineNG6, bib:GenuineNG7, bib:GenuineNG8}. To illustrate that our conclusion only refers to QNG, not simply non-Gaussianity, 
we compare the even cat state $ \mathcal{N}_\alpha^+ ( \ket{\alpha} + \ket{-\alpha} ) $ and the coherent-state mixture $ \frac{1}{2} ( \ket{\alpha}\bra{\alpha} + \ket{-\alpha}\bra{-\alpha} ) $.
Fig. 4 (a) shows resource requirement to beat NCB for those two cases. The QNG for a pure cat state increases with $\alpha$, whereas the mixed-cat state has no QNG regardless of $\alpha$. The critical squeezing for even cat states increases with QNG while that for coherent-state mixtures hardly changes. 
	\begin{figure}[t]
	\centering \includegraphics[clip=true, width=0.8\columnwidth]{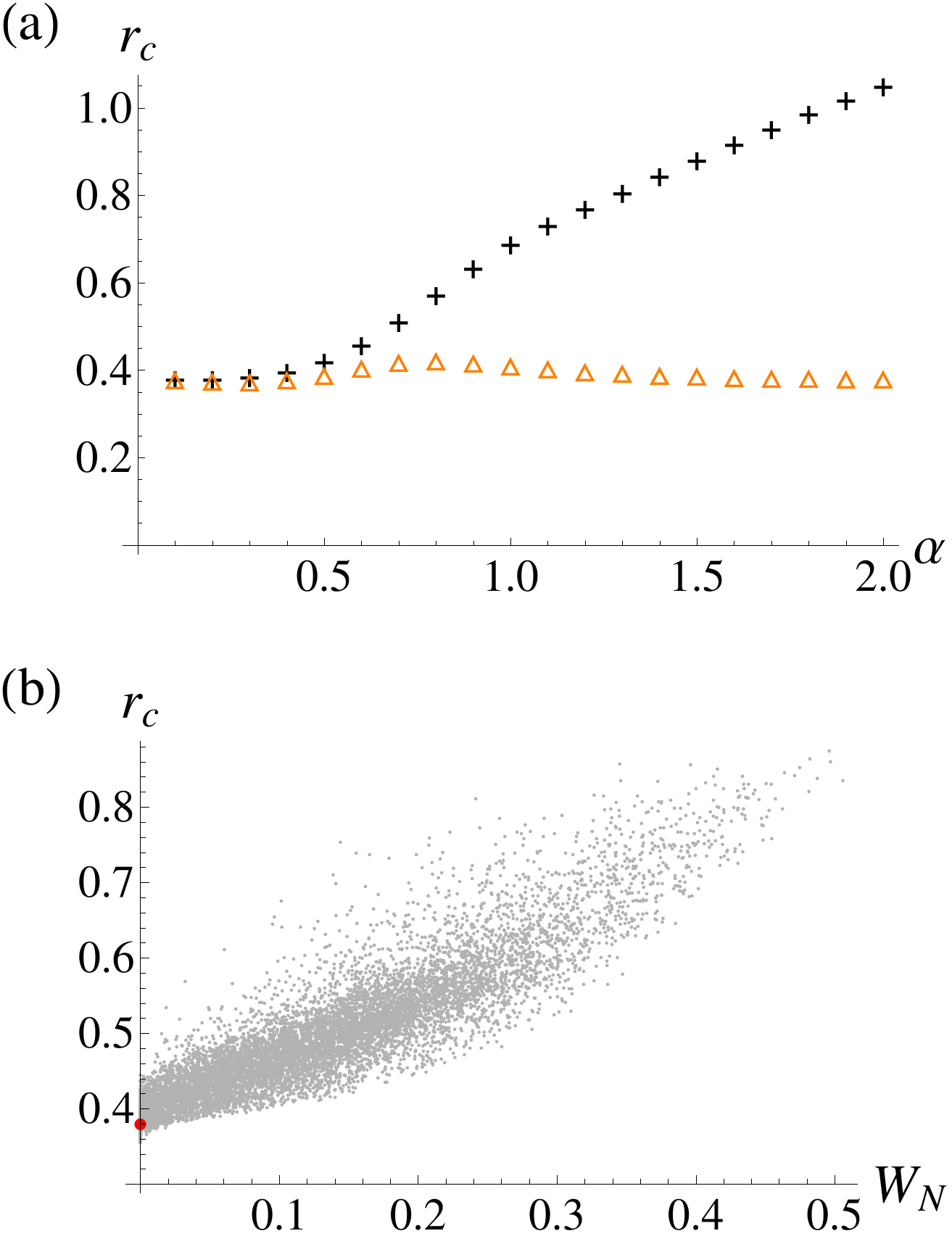}
	\caption{\label{fig:mixed} (a) Critical squeezing $r_\textrm{c}$ to achieve secure quantum teleportation with respect to $ \alpha $ for even cat states (black crosses) and for coherent-state mixtures (orange triangles). (b) critical squeezing $r_\textrm{c}$ versus logarithmic Wigner-negativity $W_\textrm{N}$. Each point corresponds to a randomly chosen mixed state in the Fock space $|0\rangle,|1\rangle$ and $|2\rangle$ (red dot: vacuum state).} 
	\end{figure}
		
 To understand the critical source of QNG more deeply, we investigate mixed input states residing in the Hilbert space spanned by $|0\rangle,|1\rangle$ and $|2\rangle$, which cover positive and non-positive Wigner functions. We then compare the logarithmic Wigner negativity $W_\textrm{N}\equiv\log[\int d^2\alpha |W(\alpha)|]$ \cite{bib:WignerNegativity1, bib:WignerNegativity2} of input state with its resource requirement $r_\textrm{c}$ to beat NCB in Fig. 4 (b). We find an overall monotonic relation between $r_\textrm{c}$ and $W_\textrm{N}$. We also note that the critical squeezing is not substantially varied among states when Wigner-negativity is absent (values along y-axis in Fig. 4 (b)). We thus draw the conclusion that QNG in terms of Wigner-negativity critically determines the resource requirement to beat NCB. This tendency is somewhat related to the point that higher Fock-state components in the resource state is needed to better teleport the non-Gaussian structure of the Wigner function. For those states with positive Wigner function, such a non-Gaussian structure is much simpler leading to a less demanding resource requirement. 

\section{Discussion}

Our study reveals that QNG is an important characteristic to consider for secure quantum communication. Overall, the NCB benchmark decreases with QNG while the resource requirement to achieve secure communication becomes more demanding with QNG. To draw our conclusion, we have extensively investigated quantum non-Gaussian states, both pure and mixed,  and remarkably found that the role of QNG is prominent when Wigner function is non-positive pointing out the Wigner-negativity as a critical factor.

While more detailed investigation, e.g. distribution of input states with a finite variance \cite{bib:PhysRevA.69.042313}, will be necessary in future, our finding already has a crucial implication for CV quantum communication. For instance, if our goal is to send information securely in the form of continuous amplitudes, our result shows that the resource requirement is minimal with coherent state inputs. So far, coherent-states have been mostly used because they are readily available CV states. Our result demonstrates they actually have a merit for secure communication in a rigorous sense. In addition, the NCB studied here can be importantly used for the security analysis of communication for a broad class of quantum non-Gaussian states. There are other interesting and critical issues to address for CV quantum informatics and we hope our work could provide a useful insight into future works.


\section{Methods}

\textbf{Optimality of covariant cloner}---
We here prove the optimality of covariant cloners by showing that, for an arbitrary cloner $ T $, one can always construct a covariant cloner $ \widetilde{T} $ resulting in the same fidelity. Let us define a cloning map $ \widetilde{T} $ by averaging over shifted cloners $ T_\beta $ of the given $T$, defined in Eq. (\ref{eq:covariantcloner}), in the whole phase space, i.e., $ \widetilde{T} (\rho) = \int d^2\beta p (\beta) T_\beta (\rho) $ with a flat distribution $ p (\beta) $. Obviously, $\widetilde{T} $ is a covariant map satisfying invariance under shifting $\widetilde{T}=\widetilde{T}_\gamma $ for all $\gamma$. The fidelity of $i$th clone in Eq. (\ref{eq:singlefidelity}) under $ \widetilde{T}$, with average over all input states $ \ket{\psi_\alpha} \equiv \hat{D}(\alpha) \ket{\psi} $, is expressed as
	\begin{align}
	& \int d^2\alpha p(\alpha) {\rm Tr} \left[ \widetilde{T} \left( \ket{\psi_\alpha}\bra{\psi_\alpha} \right) \ket{\psi_\alpha}\bra{\psi_\alpha}^{(i)} \right] \nonumber \\
	& = \int d^2\alpha d^2\beta p(\alpha) p(\beta) \nonumber \\
	& \quad \times {\rm Tr} \left[ T \left( \hat{D} (\beta) \ket{\psi_\alpha}\bra{\psi_\alpha} \hat{D}^\dagger(\beta) \right) \hat{D}_i(\beta) \ket{\psi_\alpha}\bra{\psi_\alpha}^{(i)} \hat{D}_i^\dagger (\beta) \right] \nonumber \\
	& = \int d^2\alpha d^2\beta p(\alpha) p(\beta) \nonumber \\
	& \quad \times {\rm Tr} \left[ T \left( \ket{\psi_{\alpha+\beta}}\bra{\psi_{\alpha+\beta}} \right) \ket{\psi_{\alpha+\beta}}\bra{\psi_{\alpha+\beta}}^{(i)} \right] \nonumber \\
	& = \int d^2\alpha' p(\alpha') {\rm Tr} \left[ T \left( \ket{\psi_{\alpha'}}\bra{\psi_{\alpha'}} \right) \ket{\psi_{\alpha'}}\bra{\psi_{\alpha'}}^{(i)} \right] ,
	\end{align}
which is identical with the average fidelity under a given cloner $ T $. Thus the optimal cloning fidelity can always be obtained by considering covariant cloners only.

\textbf{Gaussian invariance}---
We here demonstrate the invariance of NCB under Gaussian unitary operations $ {\hat U}_\textrm{G} $ including displacement, squeezing, and phase rotation. For any cloner $ T $ on the set of states $ \mathcal{S}_{\ket{\psi}} $, we can find a cloner $ T_{U_\textrm{G}} $ on the set of states $ \mathcal{S}_{{\hat U}_\textrm{G} \ket{\psi}} $ yielding the same average fidelity, which is given by $ T_{U_\textrm{G}} (\rho) = \hat{U}_\textrm{G}^{\otimes M} T ( \hat{U}_\textrm{G}^\dagger \rho \hat{U}_\textrm{G} ) \hat{U}_\textrm{G}^\dagger{}^{\otimes M} $. Operationally, $T_{U_\textrm{G}}$ first performs the inverse Gaussian unitary $\hat{U}_\textrm{G}^\dagger$ on an input and then applies the same cloning map $T$ followed by the Gaussian operation $\hat{U}_\textrm{G}$ on the cloned copies. 

The same fidelity under the two cloners can be shown as follows. An arbitrary ${\hat U}_\textrm{G}$ makes a linear transformation between input and output operators (Bogoliubov transformation), which yields ${\hat U}_\textrm{G}^\dag{\hat D}(\alpha){\hat U}_\textrm{G}={\hat D} ({\tilde\alpha})$, i.e. a displacement operator is transformed to another displacement operator. This gives a variant covariance property of the map  $ T_{U_\textrm{G}}$ as 
$ T_{U_\textrm{G}} \left({\hat D}(\alpha){\hat U}_\textrm{G}\rho {\hat U}_\textrm{G}^\dag{\hat D}^\dag(\alpha)\right)=({\hat D}(\alpha){\hat U}_\textrm{G})^{\otimes M}T(\rho)({\hat U}^\dag_\textrm{G}{\hat D}^\dag(\alpha))^{\otimes M}$, where we have used the covariance property of the map $T$. We then find

	\begin{align}
	& {\rm Tr} \left[ T_{U_\textrm{G}} \left({\hat D}(\alpha){\hat U}_\textrm{G}\rho {\hat U}_\textrm{G}^\dag{\hat D}^\dag(\alpha)\right) {\hat D}^{(i)}(\alpha)\hat{U}_\textrm{G}^{(i)} \rho^{(i)} \hat{U}_\textrm{G}^{\dag (i)}{\hat D}^{\dag (i)}(\alpha) \right] \nonumber \\
	& = {\rm Tr} \left[ T ( \rho) \rho^{(i)} \right],	
	\end{align}
using the invariance of trace under operator-reordering and ${\hat U}^\dag {\hat U}=I$.
That is,  finding an optimal cloner $ T $ on $ \mathcal{S}_{\ket{\psi}} $ is equivalent to finding an optimal $ T_{U_\textrm{G}} $ on $ \mathcal{S}_{{\hat U}_\textrm{G} \ket{\psi}} $. In Fig. 1 (b), the cloning scheme for $ T_{U_\textrm{G}} $ corresponds to that for $ T $ by additionally implementing $\hat{U}_\textrm{G}^\dagger$ on the input state and $\hat{U}_\textrm{G}$ on the output clones, respectively.  

\section{Acknolwedgements}
This work is supported by an NPRP grant 8-751-1-157 from Qatar National Research Fund.  J.P. acknowledges support by the National Research Foundation of Korea (NRF) grant funded by the Korea government (MSIT) (NRF-2019R1G1A1002337). The calculations were performed with the RAAD supercomputer at Texas A\&M University at Qatar.

\begin{widetext}
\vspace{1cm}
\hspace{2cm}
\setcounter{equation}{0}
\setcounter{figure}{0}
\begin{center}
{\bf Supplemental Information}
\end{center}
\section*{S1. Fidelity of the Covariant Cloner}

For a general input state $\rho_{\rm in}={\hat D}^\dag(\alpha)\rho {\hat D}(\alpha)$, the phase-space overlap between the input and the $i$th output of a covariant cloner, $F^{(i)}={\rm Tr}\left[T(\rho_{\rm in})\rho_{\rm in}^{(i)}\right] $, can be expressed in terms of characteristic functions \cite{bib:Barnett} using Eq. (3) of main text as
	\begin{align} \label{eq:singlef}
	F^{(i)} & = \frac{1}{\pi} \int d^2\xi_i \chi_\textrm{in} (-\xi_i) \chi_\textrm{out} ( 0, \cdots, 0, \xi_{i,x}, \xi_{i,p} , 0, \cdots, 0 ) \nonumber \\
	& = \frac{1}{\pi} \int d^2\xi_i \chi_\textrm{in} (-\xi_i) \chi_\textrm{in} (\xi_i) \chi_\textrm{T} \left( \Omega ( 0, \cdots, 0, \xi_{i,x}, \xi_{i,p}, 0, \cdots, 0 ) \right) \nonumber \\
	& = \frac{1}{\pi} \int d^2\xi_i |\chi_\textrm{in} (\xi_i)|^2 \tr \Big[ \rho_\textrm{T} \hat{D} (
	\overbrace{\xi_{i,p}, 0, \cdots, \xi_{i,p}, 0}^{i-1 \textrm{ copies}}, 
	\overbrace{0, \xi_{i,x}}^{i\textrm{th copy}},
	\overbrace{\xi_{i,p}, 0, \cdots, \xi_{i,p}, 0}^{M-i \textrm{ copies}}
	) \Big] \nonumber \\
	& = \left\langle \frac{1}{\pi} \int d^2\xi_i |\chi_\textrm{in} (\xi_i)|^2 \exp \left[ \sqrt{2}i \left( \xi_{i,x} \hat{x}_i - \xi_{i,p} {\textstyle \sum_{j \neq i}} \hat{p}_j \right) \right] \right\rangle_{\rho_\textrm{T}} \nonumber \\
	& = \left\langle f \left(\hat{x}_i, \textstyle{\sum_{j \neq i}} \hat{p}_j\right) \right\rangle_{\rho_\textrm{T}}, 
	\end{align}
	where 
\begin{eqnarray}
	f(w_1,w_2)\equiv\frac{1}{\pi} \int d^2\xi_i |\chi_\textrm{in} (\xi_i)|^2 e^{ \sqrt{2}i ( \xi_{i,x} w_1 - \xi_{i,p} w_2)}.
	\end{eqnarray}
 Note that the two hermitian operators, $\hat{x}_i$ and $\sum_{j \neq i} \hat{p}_j$, commute with each other, so we can simply perform c-number integration to obtain the function $f \left(\hat{x}_i,\sum_{j \neq i} \hat{p}_j\right)$ of those operators in Eq. (2). The result in Eq. (1) indicates that the fidelity of the $i$-th clone is given by the quantum average of the resource state $\rho_\textrm{T}$ over a certain positive operator $f \left(\hat{x}_i,\sum_{j \neq i} \hat{p}_j\right)$, which is determined by the characteristic function of the input state in Eq. (2).

{\bf Fock-state inputs} $|n\rangle$: Using the characteristic function $\chi_\textrm{in} (\xi)= L_n ( |\xi|^2 ) e^{ -|\xi|^2/2}$ ($L_n$: Laguerre polynomial of order $n$) and the integral relation \cite{bib:JCompApplMath.71.357} 
	\begin{equation} \label{eq:LaguerreIntegral}
	\int r dr d\theta \left[ L_n ( r^2 ) \right]^2 \exp ( -r^2 ) \exp \left[ 2ri (a \cos\theta + b \sin\theta) \right] = \pi \left[ L_n ( a^2+b^2 ) \right]^2 e^{-a^2-b^2},
	\end{equation}
we obtain the fidelity as
	\begin{align} \label{eq:singlef}
	F_n^{(i)}=  \left\langle [ L_n ( \hat{O}_i ) ]^2 e^{-\hat{O}_i} \right\rangle_{\rho_\textrm{T}} .
	\end{align} 
with  $\hat{O}_i\equiv\frac{\hat{x}_i^2+(\sum_{j \neq i} \hat{p}_j)^2}{2}$. 
To establish the no-cloning bound (NCB), we consider two output clones  $ M = 2 $, which gives the average of two fidelities as
	\begin{equation}
	\frac{1}{2} \left( F_n^{(1)}+F_n^{(2)} \right) = \frac{1}{2} \left\langle \left[ L_n \left( \frac{\hat{x}_1^2 + \hat{p}_2^2}{2} \right) \right]^2 e^{-\frac{\hat{x}_1^2 + \hat{p}_2^2}{2}} + \left[ L_n \left( \frac{\hat{x}_2^2 + \hat{p}_1^2}{2} \right) \right]^2 e^{-\frac{\hat{x}_2^2 + \hat{p}_1^2}{2}} \right\rangle_{\rho_\textrm{T}}.
	\end{equation}
Optimizing the above expression over the resource state $\rho_\textrm{T}$ gives the NCB for each case. This can be accomplished by considering the matrix elements of the operators  $\hat{f}_n^{(i)}\equiv [ L_n ( \hat{O}_i ) ]^2 e^{-\hat{O}_i}$ in Fock-state basis, using position(momentum) operator representation of Fock states, written as $ \braket{x_\theta}{j} = \frac{1}{\sqrt{\pi^{1/2} 2^n n!}} e^{-x_\theta^2/2} H_j(x_\theta) e^{-in\theta} $, with $ \hat{x}_\theta = \hat{x}\cos\theta + \hat{p}\sin\theta $ and $ H_j(x_\theta) $ being the Hermite polynomial of degree $ j $. These are given by
	\begin{align} \label{eq:fk1}
	& \bra{i,j} \hat{f}_n^{(1)} \ket{l,m} = \int dx_1 dp_2 \braket{i}{x_1} \braket{j}{p_2} \left[ L_n \left( \frac{x_1^2 + p_2^2}{2} \right) \right]^2 \exp \left( -\frac{x_1^2 + p_2^2}{2} \right) \braket{x_1}{l} \braket{p_2}{m} \nonumber \\
	& \quad = \frac{(-1)^{(i-l) /2}}{\pi^2 \sqrt{2^{i+j+l+m}i!j!l!m!}} \sum_{a,b} C_{a,b}^{(n)} \int dx_1 e^{-\frac{3}{2}x_1^2} H_i (x_1) H_l (x_1) H_a \left( \frac{x_1}{\sqrt{2}} \right) \int dp_2 e^{-\frac{3}{2}p_2^2} H_j (p_2) H_m (p_2) H_b \left( \frac{p_2}{\sqrt{2}} \right) ,
	\end{align}
where $ C_{a,b}^{(n)} $ are coefficients in Hermite series expansion of $ [ L_n ( x_1^2 + p_2^2 ) ]^2 $ given by
	\begin{equation}
	C_{a,b}^{(n)} \equiv \frac{1}{\pi 2^{a+b+1} a! b!} \int dx_1 dp_2 \exp \left( -\frac{x_1^2+p_2^2}{2} \right) H_a \left( \frac{x_1}{\sqrt{2}} \right) H_b \left( \frac{p_2}{\sqrt{2}} \right) \left[ L_n \left( \frac{x_1^2+p_2^2}{2} \right) \right]^2 .
	\end{equation}
A direct calculation leads to
	\begin{align}
	C_{a,b}^{(0)} & = \delta_{a,0} \delta_{b,0} , \nonumber \\
	\{ C_{a,b}^{(1)} \} & = \begin{pmatrix}
		1 & 0 & \frac{1}{2} & 0 & \frac{1}{16} \\
		0 & 0 & 0 & 0 & 0 \\
		\frac{1}{2} & 0 & \frac{1}{8} & 0 & 0 \\
		0 & 0 & 0 & 0 & 0 \\
		\frac{1}{16} & 0 & 0 & 0 & 0 
		\end{pmatrix} \textrm{  for } 0 \le a,b \le 4 ,
	\end{align}
and $ C_{a,b}^{(n)} $ can be obtained similarly for $ n \ge 2 $. The integrals involving three Hermite polynomials in Eq. (\ref{eq:fk1}) can be evaluated using the formula derived in Ref. \cite{bib:Hermiteintegral}, written as
	\begin{equation}
	\int dx e^{-x^2} H_a (\alpha x) H_i (\beta x) H_l (\beta x) = (-1)^{s-a} 2^{2s} \alpha^{i+l} \beta^a ~ \Gamma \left( s+\frac{1}{2} \right) {}_2F_1 \left( -i, -l, \frac{1-s}{2}; \frac{1}{2\alpha^2} \right) ~ \textrm{ when } \alpha^2 + \beta^2 = 1 ,
	\end{equation}
where $ s \equiv a+i+l $ and $ {}_2F_1$ is the Gaussian hypergeometric function. The fidelity for the second output, $ F_n^{(2)} $, can be obtained similarly by changing $ x_{1(2)} $ and $ p_{1(2)} $.

The genuine no-cloning bound can be obtained by evaluating the largest eigenvalue of $ \frac{1}{2} ( \hat{f}_n^{(1)} + \hat{f}_n^{(2)} ) $. In Fig. \ref{fig:fnctrunc}, we show numerically obtained eigenvalues in the basis of finite-photon number states $ \ket{n_1,n_2} $ with $ 0 \le n_1, n_2 \le N_\textrm{trunc} $.
	\begin{figure}[t]
	\centering \includegraphics[clip=true, width=0.4\columnwidth]{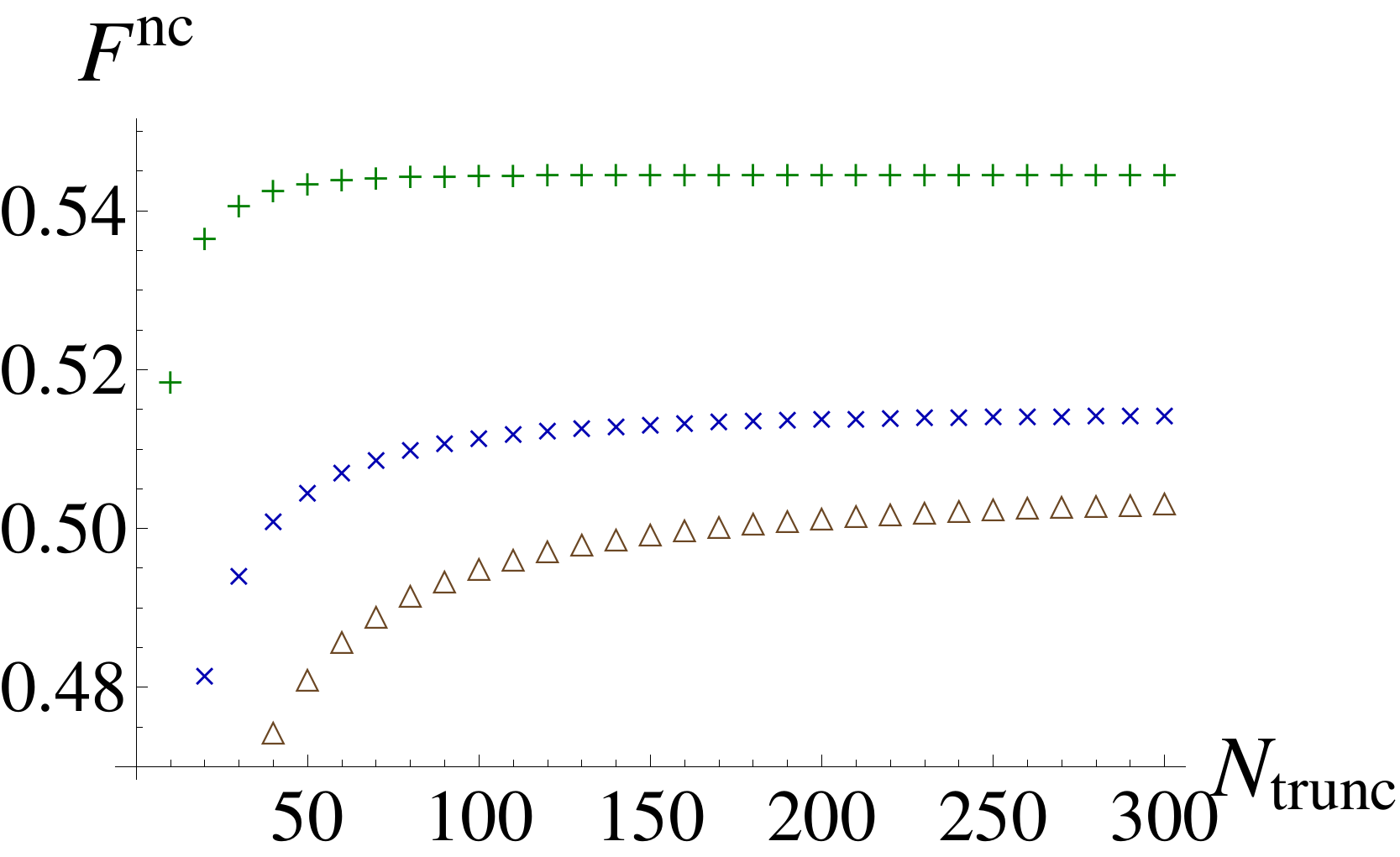}
	\caption{\label{fig:fnctrunc} No-cloning bound of DFS for $ n = $ 1(green marked +), 2(blue marked x), and 3(brown triangles) numerically calculated with different photon number truncation $ N_\textrm{trunc} $.}
	\end{figure}
It is shown that the largest eigenvalue increases as $ N_\textrm{trunc} $ increases, but becomes almost stationary at large enough $ N_\textrm{trunc} $. The difference in eigenvalues between $ N_\textrm{trunc} = 280 $ and $ N_\textrm{trunc} = 300 $ is less than $ 5 \times 10^{-7} $ for $ n = 1 $, $ 6 \times 10^{-5} $ for $ n = 2 $, and $ 3 \times 10^{-4} $ for $ n = 3 $. 

{\bf Superposition of Fock-state inputs}: A superposition of Fock states, $ c_0 \ket{0} + c_1 \ket{1} + c_2 \ket{2} $, is described by the characteristic function
	\begin{align}
	\chi_{012} ( \xi ) = \exp \left( -\frac{|\xi|^2}{2} \right) & \left[ |c_0|^2 + |c_1|^2 ( - |\xi|^2 + 1 ) + \frac{|c_2|^2}{2} ( |\xi|^4 - 4|\xi|^2 + 2 ) - ( c_0^\ast c_1 \xi^\ast - c_0 c_1^\ast \xi ) \right. \nonumber \\
	& \quad \left. + \left( \frac{c_0^\ast c_2}{\sqrt{2}} {\xi^\ast}^2 + \frac{c_0 c_2^\ast}{\sqrt{2}} \xi^2 \right) +  \left( \frac{c_1^\ast c_2}{\sqrt{2}} \xi^\ast - \frac{c_1 c_2^\ast}{\sqrt{2}} \xi \right) \left( |\xi|^2 -2 \right) \right] .
	\end{align}
We here assume that $ c_0 $ is real due to global phase and further assume that $ c_1 $ is real using the rotational invariance. Similar to Eq. (\ref{eq:singlef}), the fidelity between the input state and the 1st output state of a $ 1 \to 2 $ covariant cloner can be evaluated as
	\begin{align}
	F^{(1)} & = \left\langle \exp \left( -\frac{\hat{x}_1^2 + \hat{p}_2^2}{2} \right) \left[ c_0^4 - \frac{c_0^3 c_2}{\sqrt{2}} \left( \hat{x}_1 + i\hat{p}_2 \right)^2 - \frac{c_0^3 c_2^\ast}{\sqrt{2}} \left( \hat{x}_1 - i\hat{p}_2 \right)^2  + c_0^2 c_1^2 \left( -\hat{x}_1^2 + \hat{p}_2^2 + 2 \right) \right. \right. \nonumber \\
	& \hspace{3.5cm} - \frac{c_0^2 c_2^2}{8} \left( \hat{x}_1 + i\hat{p}_2 \right)^4 - \frac{c_0^2 {c_2^\ast}^2}{8} \left( \hat{x}_1 - i\hat{p}_2 \right)^4 + \frac{c_0^2 |c_2|^2}{2} \left( \hat{x}_1^2 + \hat{p}_2^2 - 2 \right)^2 \nonumber \\
	& \hspace{3.5cm} - \frac{c_0 c_1^2 c_2}{2\sqrt{2}} \left( \hat{x}_1^4 + \hat{p}_2^4 + 2\hat{x}_1^2 \hat{p}_2^2 - 6\hat{x}_1^2 - 2\hat{p}_2^2 - 4i\hat{x}_1 \hat{p}_2 \right) \nonumber \\
	& \hspace{3.5cm} - \frac{c_0 c_1^2 c_2^\ast}{2\sqrt{2}} \left( \hat{x}_1^4 + \hat{p}_2^4 + 2\hat{x}_1^2 \hat{p}_2^2 - 6\hat{x}_1^2 - 2\hat{p}_2^2 + 4i\hat{x}_1 \hat{p}_2 \right) \nonumber \\
	& \hspace{3.5cm} - \left( \frac{c_0 c_2^2 c_2^\ast}{8\sqrt{2}} \left( \hat{x}_1 + i\hat{p}_2 \right)^2 + \frac{c_0 c_2 {c_2^\ast}^2}{8\sqrt{2}} \left( \hat{x}_1 - i\hat{p}_2 \right)^2 \right) \left( \hat{x}_1^4 + \hat{p}_2^4 + 2\hat{x}_1^2 \hat{p}_2^2 - 8\hat{x}_1^2 - 8\hat{p}_2^2 + 8 \right) \nonumber \\
	& \hspace{3.5cm} +  \frac{c_1^4}{4} \left( \hat{x}_1^2 + \hat{p}_2^2 - 2 \right)^2 + \left( \frac{c_1^2 c_2^2}{16} \left( \hat{x}_1 + i\hat{p}_2 \right)^2 + \frac{c_1^2 {c_2^\ast}^2}{16} \left( \hat{x}_1 -i\hat{p}_2 \right)^2 \right) \left( \hat{x}_1^2 + \hat{p}_2^2 - 4 \right)^2 \nonumber \\
	& \hspace{3.5cm} + \frac{c_1^2 |c_2|^2}{4} \left( \hat{x}_1^4 + \hat{p}_2^4 + 2\hat{x}_1^2 \hat{p}_2^2 - 4\hat{x}_1^2 - 4\hat{p}_2^2 + 8 \right) \nonumber \\
	& \hspace{3.5cm} \left. \left. + \frac{|c_2|^4}{64} \left( \hat{x}_1^4 + \hat{p}_2^4 + 2\hat{x}_1^2 \hat{p}_2^2 - 8\hat{x}_1^2 - 8\hat{p}_2^2 + 8 \right)^2 \right] \right\rangle_{\rho_\textrm{T}} .
	\end{align}
The fidelity of the second output can be obtained by changing $ \hat{x}_{1(2)} $ and $ \hat{p}_{1(2)} $.

{\bf Cat-state inputs}: A cat state has the form $ \mathcal{N}_{\alpha,\gamma}^2 \big[ \ket{\alpha}\bra{\alpha} + \ket{-\alpha}\bra{-\alpha} + \gamma \left( \ket{\alpha}\bra{-\alpha} + \ket{-\alpha}\bra{\alpha} \right) \big] $, where $ \mathcal{N}_{\alpha,\gamma} = \left( 2 + 2 \gamma e^{-2 \alpha^2} \right)^{-1/2} $ is the normalization factor and we assume $ \alpha $ is real. When $ \gamma = 1 (-1) $, the state represents even(odd) cat state and when $ \gamma = 0 $, it becomes a mixture of coherent states. Its characteristic function is written as
	\begin{equation}
	\chi_\textrm{cat} ( \xi ) = 2\mathcal{N}_{\alpha,\gamma}^2 \exp \left( -\frac{|\xi|^2}{2} \right) \left[ \cos \left( 2 \alpha \xi_p \right)  + \gamma e^{-2 \alpha^2} \textrm{cosh} \left( 2 \alpha \xi_x \right) \right] .
	\end{equation}
The fidelity of cloner can be evaluated as
	\begin{align}
	F^{(1)} & =  \left\langle 2{\mathcal{N}_{\alpha,\gamma}}^4 \exp \left( -\frac{\hat{x}_1^2 + \hat{p}_2^2}{2} \right) \left[ 1 + e^{-4\alpha^2} \textrm{cosh} \left( 2\sqrt{2} \alpha \hat{p}_2 \right) + 2\gamma  e^{-2\alpha^2} \cos \left( \sqrt{2} \alpha \hat{x}_1 \right) \textrm{cosh} \left( \sqrt{2} \alpha \hat{p}_2 \right) \right. \right. \nonumber \\
	& \hspace{4.5cm} \left. \left. + \gamma^2 e^{-4\alpha^2} + \gamma^2 \cos \left( 2\sqrt{2} \alpha \hat{x}_1 \right) \right] \right\rangle_{\rho_\textrm{T}} .
	\end{align}
	
\section*{S2. Ansatz solution and iteration method for NCB of Fock-state inputs}

In this section, we take a different, more intuitive, approach to investigate the NCB for the Fock-state inputs. On examining the structure of Eq. (5) to determine NCB, we see that it involves only two-photon processes $\{\hat{x}_i^2, \hat{p}_i^2\}$ and is also symmetric under the permutation of two modes. A plausible optimal solution is then given by $|\Psi\rangle_{12}\sim |\phi\rangle_1|\phi'\rangle_1+|\phi'\rangle_1|\phi\rangle_2$
where $|\phi\rangle\equiv\sum_mC_{2m}|2m\rangle$ is a state involving only even numbers and $|\phi'\rangle\equiv e^{i\frac{\pi}{2}\hat{n}}|\phi\rangle=\sum_m(-1)^{2m}C_{2m}|2m\rangle$ is the state obtained by acting $\frac{\pi}{2}$ phase-shift on $|\phi\rangle$.  

Before we proceed, as a remark, if we treat the observables ${\hat z}_1\equiv \frac{\hat{x}_1^2 + \hat{p}_2^2}{2}$ and  ${\hat z}_2\equiv \frac{\hat{x}_2^2 + \hat{p}_1^2}{2}$ as classical variables ignoring their commutation relation, we immediately see that the optimal cloning fidelity would be 1 regardless of $n$ due to $f_n^{(1)}(z_1)=f_n^{(2)}(z_2)=1$ at $z_1=z_2=0$ (See the profiles of the function $f_n(z) = [L_n(z)]^2 e^{-z}$ in Fig. 2 (a)). This is of course not achievable due to uncertainty principle as both of the probability distributions $P(z_1)$ and $P(z_2)$ corresponding to the two noncommuting observables cannot be a delta function. In this respect, optimizing the cloning bound is equivalent to finding narrow-enough distributions allowed by uncertainty principle, in particular the symmetric ones $P(z_1)=P(z_2)$. 

\subsection{Aansatz: two-mode superposition of squeezed states}

Let us first consider a squeezed state $ \ket{\Phi_r}_{12} = \hat{S}_1 (r) \otimes \hat{S}_2 (-r) \ket{00}_{12} $, where $ \hat{S}_i\equiv e^{\frac{r}{2}(a_i^2-a_i^{\dag2})} $ ($i=1,2$) is a single-mode squeezing operation along $\hat x$-quadrature. This gives the highest fidelity $ f^{(1)} = 1 $ of the 1st clone in the limit of infinite squeezing $s\rightarrow\infty$ with the probability distribution $P(z_1)\sim\delta(z_1)$, while it makes the fidelity of the 2nd clone worst. A similar argument goes with the squeezed state $ \ket{\Phi'_r}_{12} = \hat{S}_1 (-r) \otimes \hat{S}_2 (r) \ket{00}_{12} $ along orthogonal directions. Following the above argument, we instead come up with a two-mode superposition of squeezed states, i.e. $ \ket{\Phi_r}_{12}+\ket{\Phi'_r}_{12}$ as an ansatz, which can be written in terms of quadrature eigenstates $ \ket{x_1} $ and $ \ket{p_2} $ as
	\begin{equation} \label{eq:sqsup}
	\mathcal{N}_r \int dx_1 dp_2 \left[ \frac{1}{\sqrt{\pi e^{-2r}}} \exp\left( -\frac{1}{2e^{-2r}} ( x_1^2 + p_2^2 ) \right) + \frac{1}{\sqrt{\pi e^{2r}}} \exp\left( -\frac{1}{2e^{2r}} ( x_1^2 + p_2^2 ) \right) \right] \ket{x_1} \ket{p_2} ,
	\end{equation}
where $ \mathcal{N}_r = ( 2 + 2 \textrm{sech} (2r) )^{-1/2} $ is the normalization factor. Then we have the analytic form of fidelity as a function of squeezing $r$ given by
	\begin{align} \label{eq:fsqsup}
	F_n^{(1)}(r) = F_n^{(2)}(r) & = \mathcal{N}_r^2 \int dx_1 dp_2 \left[ L_n \left( \frac{x_1^2+p_2^2}{2} \right) \right]^2 e^{-\frac{x_1^2+p_2^2}{2}} \nonumber \\
	& \hspace{1.5cm} \times \left[ \frac{1}{\sqrt{\pi e^{-2r}}} \exp\left( -\frac{1}{2e^{-2r}} ( x_1^2 + p_2^2 ) \right) + \frac{1}{\sqrt{\pi e^{2r}}} \exp\left( -\frac{1}{2e^{2r}} ( x_1^2 + p_2^2 ) \right) \right]^2 \nonumber \\
	& = \frac{2 \mathcal{N}_r^2 \Gamma \left(n+\frac{1}{2}\right)}{\sqrt{\pi } n!} \left[ \frac{\, _2F_1\left(\frac{1}{2},-n;\frac{1}{2}-n;\left(1-\frac{2}{1+2 e^{-2 r}}\right)^2\right)}{e^{2 r}+2} + \frac{\, _2F_1\left(\frac{1}{2},-n;\frac{1}{2}-n;\left(1-\frac{2}{1+2 e^{2 r}}\right)^2\right)}{e^{-2 r}+2} \right. \nonumber \\
	& \hspace{2.5cm} \left. +\frac{2 \, _2F_1\left(\frac{1}{2},-n;\frac{1}{2}-n;\left(1-\frac{2}{1+e^{-2 r}+e^{2 r}}\right)^2\right)}{e^{-2 r}+e^{2 r}+1} \right] ,
	\end{align}
with $ \Gamma(n) $ the Gamma function and $  _2F_1(a,b;c;z) $ the hypergeometric function. For each $n$, we can find an optimal squeezing $r$ to maximize $F_n^{(1)}(r)=F_n^{(2)}(r)$.  

In Fig. \ref{fig:fsqsup} (b), we show the cloning fidelity with respect to the squeezing parameter for the case of $ n = 1 $, in comparison with our numerical NCB obtained in Fock-state basis. We observe that the cloning fidelity is maximized with a finite squeezing and that the maximized fidelity is close to but slightly lower than our numerical NCB. In Fig. \ref{fig:fsqsup}(c), we show the fidelity optimized over the squeezing $r$ for each $ n=0,1,2,3$, which is close to the NCB numerically obtained in Fock-state basis. This suggests that the superposition of squeezed states $ \ket{\Phi_r}_{12}+\ket{\Phi'_r}_{12}$ can be employed as a resource for cloning close to optimal one.
	\begin{figure}[t]
	\centering \includegraphics[clip=true, width=0.8\columnwidth]{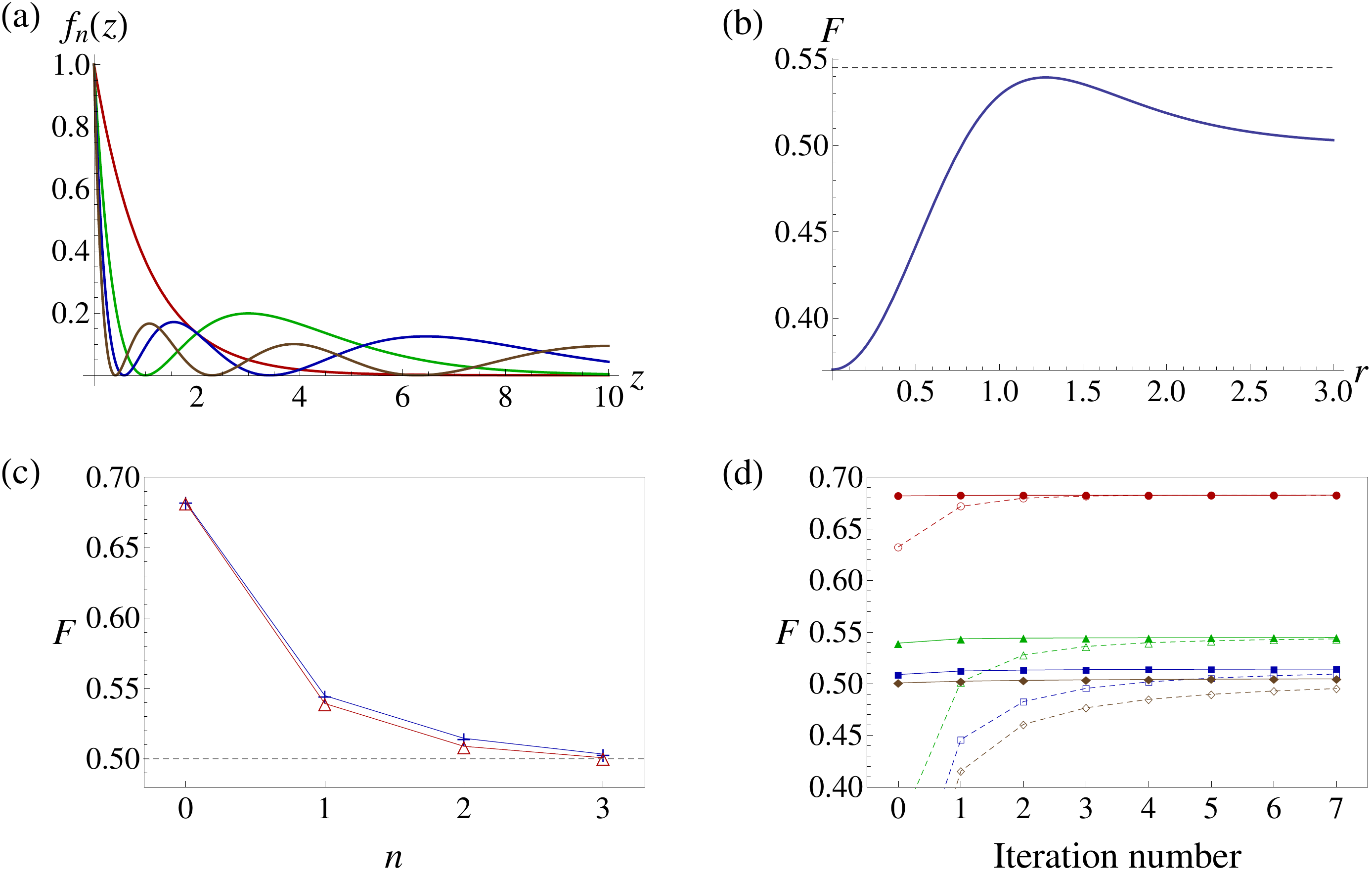}
	\caption{\label{fig:fsqsup} (a) Profiles of the functions $f_n(z) = [L_n(z)]^2 e^{-z}$ for $ n = $  0(red), 1(green), 2(blue), 3(brown). (b) cloning fidelity (\ref{eq:fsqsup}) against squeezing parameter $r $ for the case of $ |n=1\rangle $. Horizontal dashed line represents the NCB obtained in Fock-state basis. (c) Red: cloning fidelity (\ref{eq:fsqsup}) using two-mode superposition of squeezed states optimized over squeezing parameter $r$ for each case of $n$. Blue: NCB obtained in Fock-state basis. (d) Solid lines: cloning fidelity obtained by iteration method on the two-mode superposition of squeezed states for $ n = $ 0(red), 1(green), 2(blue), 3(brown). Dashed lines: cloning fidelity obtained by iteration method on the vacuum-state. }
	\end{figure}
	
\subsection{Numerical iteration method}

The previously identified superposition of squeezed states gives the cloning fidelity close to the ultimate NCB but there exists a gap between two quantities, which slightly increases with $n$. We may further optimize the fidelity based on an iteration method: If a positive Hermitian operator $\hat P$ is expressed using its eigen-state expansion as ${\hat P}=\sum_jp_j|j\rangle\langle j|$ with eigenvalues in decreasing order, i.e. $p_0\ge p_1\ge p_2\cdots$, the action of the operator $\hat P$ multiple times gives ${\hat P}^M=\sum_jp_j^M|j\rangle\langle j|\rightarrow p_0^M|0\rangle\langle0|$ as $M$ grows. That is, acting $\hat P$ many times essentially projects a given state to its maximum eigenstate while its convergence may depend on the initial state.  

We apply this iteration method to the superposition of squeezed state above, which is already close to the optimal solution.  In each step of iteration, we evaluate the subsequent wavefunction by applying the operator $ \frac{1}{2} ( \hat{f}_n^{(1)} + \hat{f}_n^{(2)} ) $ on the wavefunction. We begin with the wavefunction $ \ket{\Psi_0} = \int dx_1 dp_2 \psi_0 ( x_1, p_2 ) \ket{x_1,p_2} $, where $ \psi_0 ( x_1, p_2 ) = \frac{\mathcal{N}_r}{\sqrt{2\pi}} \left[ \exp\left( -\frac{1}{2} ( e^{2r} x_1^2 + e^{-2r} p_2^2 ) \right) + \exp\left( -\frac{1}{2} ( e^{-2r} x_1^2 + e^{2r} p_2^2 ) \right) \right] $. The subsequent wavefunction can be evaluated as
	\begin{align}
	 \ket{\tilde{\Psi}_{i+1}} & = \int dx_1 dp_2 \tilde{\psi}_{i+1} ( x_1, p_2 ) \ket{x_1,p_2} = \frac{1}{2} \left( \hat{f}_n^{(1)} + \hat{f}_n^{(2)} \right) \int dx_1 dp_2 \psi_i ( x_1, p_2 ) \ket{x_1,p_2} , \\
	 \tilde{\psi}_{i+1} ( x_1, p_2 ) & = \frac{1}{2} \left[ \psi_i ( x_1, p_2 ) \left[ L_n \left( \frac{x_1^2 + p_2^2}{2} \right) \right]^2 \exp \left( -\frac{x_1^2 + p_2^2}{2} \right) \right. \nonumber \\
	 & \hspace{1cm} + \frac{1}{2\pi} \int dx_2 dp_1 e^{i ( x_1 p_1 - x_2 p_2 )} \left[ L_n \left( \frac{x_2^2 + p_1^2}{2} \right) \right]^2 \exp \left( -\frac{x_2^2 + p_1^2}{2} \right) \nonumber \\
	  & \hspace{3.5cm} \left. \times \left( \frac{1}{2\pi} \int dx_1' dp_2' e^{i ( - x_1' p_1 + x_2 p_2' )} \psi_i ( x_1', p_2' ) \right) \right] . \nonumber
	\end{align}
followed by a proper normalization $ \ket{\Psi_{i+1}} = \ket{\tilde{\Psi}_{i+1}} / \big\lVert \ket{\tilde{\Psi}_{i+1}} \big\rVert $ .
After 7 steps of iteration, we obtain the fidelity $ 0.5447 $ for $ n = 1 $ and $ 0.5141 $ for $ n = 2 $. Compared to the NCB obtained in Fock basis, we have identical results with the precision to three decimal places, and the iteration method gradually approaches the NCB in Fock-state basis from below.	

In Ref. \cite{bib:Kruger}, the iteration method was also used to find the NCB of coherent states, where the trial wavefunction was assumed to be a Gaussian vacuum state. For comparison, we also adopt the vacuum state under iteration in Fig. 2 (d), which exhibits a much slower increase of fidelity than our superposition of squeezed states.  

\subsection{ Profiles of optimal solutions}
We here look into the probability distribution $P(z_1)$ of the observable ${\hat z}_1\equiv\frac{{\hat x}_1^2+{\hat p}_2^2}{2}$ for near-optimal and optimal solutions in comparison with the profile of the test function $f_n(z) = [L_n(z)]^2 e^{-z}$. Note that the probability $P(z_2)$ of the observable ${\hat z}_2\equiv\frac{{\hat x}_2^2+{\hat p}_1^2}{2}$ must be identical with $P(z_1)$ due to the symmetry of the solutions. The cloning fidelity is thus expressed as $\int_0^\infty dz P(z)f_n(z)$, that is, the overlap between $P(z)$ and $f_n(z)$ determines the output fidelity.

	\begin{figure}[t]
	\centering \includegraphics[clip=true, width=\columnwidth]{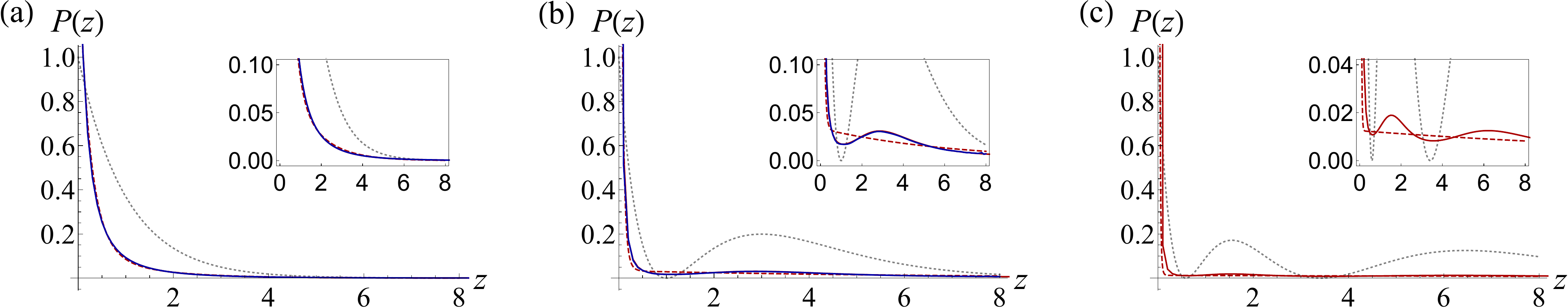}
	\caption{\label{fig:pzprofile} Illustration of $ P(z_1) $ and $ f_n(z) $ for (a) $ n = 0 $, (b) $ n = 1 $, and (a) $ n = 2 $. Gray dotted curves represent $ f_n(z) $. Red solid (dashed) curves correspond to the wavefunction after (before) 7-iterations, starting with the superpostion of squeezed states. Blue solid curves correspond to the optimal state obtained in Fock-state basis. Red and blue solid curves are almost indiscernible.}
	\end{figure}
In Fig. \ref{fig:pzprofile} (a)-(c), we show $P(z_1)$ and $f_n(z)$ for the cloning of each Fock state $|n\rangle$ ($n=0,1,2$). As mentioned before, if $P(z_1)$ is a delta-function, the maximal fidelity $F_n$ would achieve 1 regardless of $n$ since $f_n(z=0)=1$ at $z=0$. Due to uncertainty principle and the symmetry condition $P(z_1)=P(z_2)$ required for the optimal cloning, $P(z_1)$ cannot be sharply peaked indefinitely. Instead, as shown in the figures, $P(z_1)$ rougly takes after the shape of the function $f_n(z)$ to maximize fidelity. For instance, the optimal distribution $P(z_1)$ for $n=1$ slightly exploits the second peak around $z=3$ of $f_1(z)$ rather than becoming a simple decaying distribution from $z=0$.

 This picture also provides an insight into why NCB decreases with respect to $n$. For instance, comparing two functions $f_0(z)=e^{-z}$ and $f_1(z) = [L_1(z)]^2 e^{-z}$, we see that $f_0(z)$ is overall larger than $f_1(z)$ except the region around $z=3$ (second maximum of $f_1(z)$ in Fig. 2 (a)). Therefore, with the probability distribution $P(z)$ allowed by uncertainty principle, $F_0=\int_0^\infty dz P(z)f_0(z)$ must be larger than $F_1=\int_0^\infty dz P(z)f_1(z)$. Note that the value of the second maximum in $f_1(z)$ is as small as 0.2, so this region cannot be exploited to beat the NCB of $n=0$, which is as large as 0.6826.
 
\section*{S3. $ 1 \to M $ Symmetric Gaussian Cloner}

As explained in the main text, the covariant cloning map $ T $ is described, in the characteristic function formalism, as
	\begin{equation} \label{eq:covariantcloner}
	\chi_\textrm{out} ( \boldsymbol{\xi} ) = \chi_\textrm{T} ( \Omega \boldsymbol{\xi} ) \chi_\textrm{in} \left( \textstyle{\sum_i} \xi_i \right) \equiv t ( \boldsymbol{\xi} ) \chi_\textrm{in} \left( \textstyle{\sum_i} \xi_i \right) .
	\end{equation}
A Gaussian cloner is characterized by a Gaussian function $ t ( \boldsymbol{\xi} ) = \exp \left( - \frac{1}{2} \boldsymbol{\xi}^\mathrm{T} \cdot \gamma_t \cdot \boldsymbol{\xi} \right) $ where $ \gamma_t $ is the covariance matrix of the function $ t ( \boldsymbol{\xi} ) $ \cite{bib:GQI}. Note that $ t ( \boldsymbol{\xi} ) $ itself is not a characteristic function of a physical state due to the transformation $\Omega$ used in $\chi_\textrm{T} ( \Omega \boldsymbol{\xi} )$ (main text). We here assume a symmetric cloner, that is, the map (\ref{eq:covariantcloner}) is symmetric under permutation of modes. It is fulfilled with a symmetric covariance matrix of the form
	\begin{equation} \label{eq:transformationCM}
	\gamma_t = \left( \begin{array}{ccccc}
		A & B & B & & B \\
		B & A & B & \cdots & B \\
		B & B & A & & B \\
		& \vdots & & \ddots & \vdots \\
		B & B & B & \cdots & A
	\end{array} \right) ,
	\end{equation}
where $ A $ and $ B $ represent $ 2 \times 2 $ matrices. Furthermore, the map (\ref{eq:covariantcloner}) has invariance under phase rotation, that is, $ \chi_\textrm{out} ( \boldsymbol{\xi} e^{i\phi} ) = t ( \boldsymbol{\xi} ) \chi_\textrm{in} \left( \sum_i \xi_i e^{i\phi} \right) $. The rotational invariance is satisfied if and only if $ A = a \mathbb{1}_2 $ and $ B = b \mathbb{1}_2 $, where $ \mathbb{1}_d $ is a $ d \times d $ identity matrix.

The transformation $ T $ is positive when output states are always positive for any positive input states. An $ M $-mode state is positive if the covariance matrix $ \gamma $ obeys
	\begin{equation}
	\gamma + i \sigma_M \ge 0 ~ , \textrm{ where } \sigma_M = \mathbb{1}_M \otimes \left( \begin{array}{cc}
		0 & 1 \\
		-1 & 0 
	\end{array} \right) .
	\end{equation}
Then, the transformation (\ref{eq:covariantcloner}) is positive if \cite{bib:Kruger}
	\begin{equation}
	\gamma_t + i ( \sigma_M - \mathbb{E}_M \otimes \sigma_1 ) \ge 0 ,
	\end{equation}
where $ \mathbb{E}_d $ is a $ d \times d $ matrix with $ (\mathbb{E}_d )_{i,j} = 1 $ for all $ i, j $. With Eq. (\ref{eq:transformationCM}), the positivity condition reads $ a-b \ge 1 $ and $ a + (M-1)b \ge M-1 $.

From Eq. (\ref{eq:singlef}), the single-copy fidelity is written as
	\begin{align}
	f^{(1)} & = \frac{1}{\pi} \int d\xi_1^2 \left[ L_n ( |\xi_1|^2) \right]^2 \exp \left( - \frac{a+2}{2} |\xi_1|^2 \right) \nonumber \\
	& = \begin{dcases}
		\frac{2}{a+2}, & n = 0 , \\
		\frac{2a^2+8}{(a+2)^3}, & n= 1 , \\
		\frac{2a^4+32a^2+512}{(a+2)^5}, & n = 2 , \\
		\frac{2a^6+64a^4+512a^4+2048}{(a+2)^7}, & n = 3 .
		\end{dcases}
	\end{align}
As we consider a symmetric cloner, fidelity of each output state is the same so that it suffices to maximize $ f^{(1)} $. We find the optimal fidelity of symmetric Gaussian cloner given by
	\begin{align}
	F_0^{\textrm{(G)}1 \to M} & = \frac{M}{2M-1} , \nonumber \\
	F_1^{\textrm{(G)}1 \to M} & = \frac{M(2M^2-2M+1)}{(2M-1)^3} , \\ \nonumber
	F_2^{\textrm{(G)}1 \to M} & = \frac{M(6M^4-12M^3+10M^2-4M+1)}{(2M-1)^5} \\ \nonumber
	F_3^{\textrm{(G)}1 \to M} & = \frac{M(2M^2-2M+1)(10M^4-20M^3+14M^2-4M+1)}{(2M-1)^7} .
	\end{align}
Although the fidelity shows different values for different input states, the maximum is achieved at the same condition: $ a = \frac{2M-2}{M} $ and $ b = \frac{M-2}{M} $. This means that the optimal $ 1 \to M $ Gaussian cloner for the coherent-state input is also optimal for the displaced Fock-state input when restricted to Gaussian protocols. The case of interest is $ M = 2 $, which yields the Gaussian no-cloning bound: $ F_0^\textrm{(G)nc} = \frac{2}{3} \simeq 0.6667 $, $ F_1^\textrm{(G)nc} = \frac{10}{27} \simeq 0.3704 $, $ F_2^\textrm{(G)nc} = \frac{22}{81} \simeq 0.2716 $, and $ F_3^\textrm{(G)nc} = \frac{490}{2187} \simeq 0.2241 $.

Note that, when $ M \to \infty $, the cloner behaves as a classical measure-and-prepare transformation employing Gaussian measurement and Gaussian operations only, leading to fidelities $ \frac{1}{2} $, $ \frac{1}{4} $, $ \frac{3}{16} $, and $ \frac{5}{32} $, respectively for $ n = 0, 1, 2, 3 $. We find that those values are indeed the genuine classical bound even though non-Gaussian operations are allowed, which is shown in the following Section. This means that the classical bound can be achieved with a Gaussian scheme, specifically, balanced heterodyne detection and coherent-state preparation.

\section*{S4. Classical Bound}


Let us consider a classical cloning transformation $ T $ assuming a covariant cloner. Since $ T $ is a classical operation, it has to be positive under time reversal. This is fulfilled if the transformation by $ T $ is described as
	\begin{equation}
	\chi_\textrm{out} (\xi) = \chi_\textrm{T} (\sqrt{2}\xi) \chi_\textrm{in} (\xi) ,
	\end{equation}
where $ \chi_\textrm{T} (\xi) $ is the characteristic function of a quantum state $ \rho_\textrm{T} $ \cite{bib:Kruger}. The classical bound is obtained when the fidelity is maximized over all quantum state $ \rho_\textrm{T} $, that is,
	\begin{align}
	F_n^\textrm{cl} & = \max_{\rho_\textrm{T}} \frac{1}{\pi} \int d^2\xi \chi_\textrm{in} (-\xi) \chi_\textrm{out} (\xi) \nonumber \\
	& = \max_{\rho_\textrm{T}} \frac{1}{\pi} \int d^2\xi \left[ L_n ( |\xi|^2 ) \right]^2 \exp ( -|\xi|^2 ) \chi_\textrm{T} (\sqrt{2}\xi) \nonumber \\
	& = \max_{\rho_\textrm{T}} \frac{1}{\pi} \int \frac{d^2\xi'}{2} \left[ L_n \left( \frac{|\xi'|^2}{2} \right) \right]^2 \exp \left( -\frac{|\xi'|^2}{2} \right) \chi_\textrm{T} (\xi') \nonumber \\
	& = \max_{\rho_\textrm{T}} \tr \left[ \hat{A}_n \rho_\textrm{T} \right] , \nonumber \\
	\end{align}
where $ \hat{A}_n $ is the operator described by the characteristic function
	\begin{align}
	\chi_{A_n} ( \xi ) \equiv \tr \left[ \hat{A}_n \hat{D} (\xi) \right] = \frac{1}{2} \left[ L_n \left( \frac{|\xi|^2}{2} \right) \right]^2 \exp \left( -\frac{|\xi|^2}{2} \right) .
	\end{align}
Since Fock states have the characteristic function of the form $ \chi_{n} = L_n ( |\xi|^2 ) \exp ( -|\xi|^2/2 ) $, $ \hat{A}_n $ can be directly obtained by the Laguerre series expansion of $ \frac{1}{2} \left[ L_n \left( \frac{|\xi|^2}{2} \right) \right]^2 $, which leads to
	\begin{align}
	\hat{A}_0 & = \frac{1}{2}\ket{0}\bra{0} , \nonumber \\
	\hat{A}_1 & = \frac{1}{4}\ket{0}\bra{0} + \frac{1}{4}\ket{2}\bra{2} , \nonumber \\
	\hat{A}_2 & = \frac{3}{16}\ket{0}\bra{0} + \frac{1}{8}\ket{2}\bra{2} + \frac{3}{16}\ket{4}\bra{4} , \nonumber \\
	\hat{A}_3 & = \frac{5}{32}\ket{0}\bra{0} + \frac{3}{32}\ket{2}\bra{2} + \frac{3}{32}\ket{4}\bra{4} + \frac{5}{32}\ket{6}\bra{6} .
	\end{align}
Therefore, we obtain the classical bound $ F_0^\textrm{cl} = \frac{1}{2} $, $ F_1^\textrm{cl} = \frac{1}{4} $, $ F_2^\textrm{cl} = \frac{3}{16} $, and $ F_3^\textrm{cl} = \frac{5}{32} $. and the bound is achieved with $ \rho_\textrm{T} = \ket{0}\bra{0} $ or $ \ket{2n}\bra{2n} $. By choosing $ \rho_\textrm{T} = \ket{0}\bra{0} $, we find that the classical bound can be achieved with Gaussian schemes.

 \section*{S5. Plot of resource requirement for various pure input states}
  We here provide the plot showing the resource requirement $r_\textrm{c}$ against the Wigner negativity $W_\textrm{N}\equiv\log[\int d^2\alpha |W(\alpha)|]$ of input state, which corresponds to Fig. 3 of main text. On comparing these two figures, we see that the overall trend is the same, i.e., $r_\textrm{c}$ becomes larger with QNG whether the latter is quantified in terms of relative entropy $\delta$ or the wigner negativity $W_\textrm{N}$. 
 
 \begin{figure}[t]
	\centering \includegraphics[clip=true, width=0.5\columnwidth]{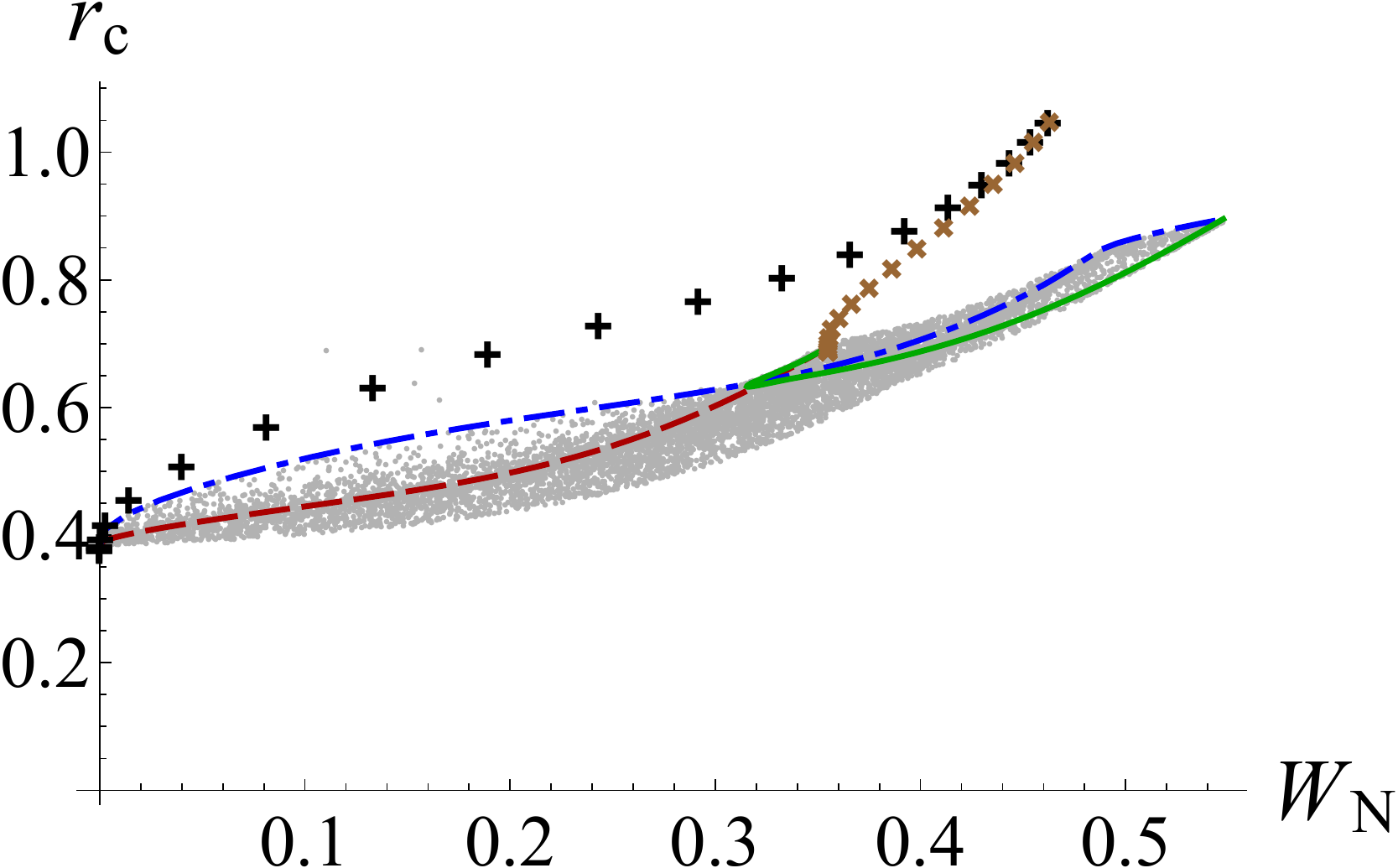}
	\caption{\label{fig:WN} Wigner negativity $W_\textrm{N}$ versus critical squeezing $r_\textrm{c}$ to achieve secure quantum teleportation for the same quantum non-Gaussian states as those in Fig. 3 of main text. }
	\end{figure}
	
\section*{S6. Optimal teleportation fidelity under energy constraint}


In this section, we investigate the optimal teleportation fidelity for DFS input states. In our previous work \cite{bib:PhysRevA.95.052343}, we investigated the optimal teleportation fidelity under energy constraint for the coherent-state input, and we here follow the same method for DFS input. Using the characteristic function of DFS, $ \chi_{n, \alpha} (\xi) =  L_n ( |\xi|^2 ) \exp ( -|\xi|^2/2 + \xi \alpha^\ast - \xi^\ast \alpha ) $, the teleportation fidelity is written as
	\begin{align} \label{eq:fidelity}
	F_n & = \frac{1}{\pi} \int d^2\xi \chi_\textrm{in} (-\xi) \chi_\textrm{out} (\xi) \nonumber \\
	& = \frac{1}{\pi} \int d^2\xi \chi_{n, \alpha} (-\xi) \chi_{n, \alpha} (\xi) \chi_\textrm{AB} (\xi^\ast, \xi) \nonumber \\
	& = \frac{1}{\pi} \int d^2\xi d^2\eta \left[ L_n ( |\xi|^2 ) \right]^2 \exp ( -|\xi|^2 ) \chi_\textrm{AB} (\xi, \eta) \delta ( \xi^\ast - \eta ) ,
	\end{align}
where $ \delta (x) $ is the Dirac delta function. On the other hand, if $ F_n $ can be evaluated as an expectation value of the fidelity operator $ \hat{\mathcal{F}}_n $, it can be written as
	\begin{equation} \label{eq:operatorrep}
	F_n = \frac{1}{\pi^2} \int d^2\xi d^2\eta ~ \chi_{F_n} (\xi, \eta) \chi_\textrm{AB} (\xi, \eta) ,
	\end{equation}
where $ \chi_{F_n} (\xi, \eta) $ represents the characteristic function of $ \hat{\mathcal{F}}_n $. By comparing Eqs. (\ref{eq:fidelity}) and (\ref{eq:operatorrep}), we find
	\begin{equation}
	\chi_{F_n} (\xi, \eta) = \pi \left[ L_n ( |\xi|^2 ) \right]^2 \exp ( -|\xi|^2 ) \delta ( \xi^\ast - \eta ) .
	\end{equation}
Then the fidelity operator becomes
	\begin{align}
	\hat{\mathcal{F}}_n & = \frac{1}{\pi^2} \int d^2\xi d^2\eta ~ \chi_{F_n} (\xi, \eta) \hat{D}_\textrm{A}^\dagger (\xi) \hat{D}_\textrm{B}^\dagger (\eta) \nonumber \\
	& = \frac{1}{\pi} \int d^2\xi \left[ L_n ( |\xi|^2 ) \right]^2 \exp ( -|\xi|^2 ) \hat{D}_\textrm{A}^\dagger (\xi) \hat{D}_\textrm{B}^\dagger (\xi^\ast) \nonumber \\
	& = \frac{1}{\pi} \int r dr d\theta \left[ L_n ( r^2 ) \right]^2 \exp ( -r^2 ) \exp \left[ 2ir ( \cos\theta \hat{v} + \sin\theta \hat{u} ) \right] ,
	\end{align}
where $ \xi = r e^{i \theta} $. Since the operators $ \hat{u} = \frac{1}{\sqrt{2}} ( \hat{x}_\textrm{A} - \hat{x}_\textrm{B} ) $ and $ \hat{v} = \frac{1}{\sqrt{2}} ( \hat{p}_\textrm{A} + \hat{p}_\textrm{B} ) $ commute with each other, they can be considered as independent variables during integration.
We again refer to Eq. (\ref{eq:LaguerreIntegral}) to evaluate the integration and obtain
	\begin{equation} \label{eq:Foperator}
	F_n = \left\langle \hat{\mathcal{F}}_n \right\rangle_{\rho_\textrm{AB}}
	= \left\langle [ L_n ( \hat{O}_\textrm{EPR} ) ]^2 e^{-\hat{O}_\textrm{EPR}} \right\rangle_{\rho_\textrm{AB}} ,
	\end{equation}
where the EPR operator is given by $\hat{O}_\textrm{EPR}\equiv \hat{u}^2+ \hat{v}^2=\frac{1}{2}( \hat{x}_\textrm{A} - \hat{x}_\textrm{B} )^2+\frac{1}{2} ( \hat{p}_\textrm{A} + \hat{p}_\textrm{B} )^2$.  That is, under the covariant BK teleportation scheme, the output fidelity is solely determined by the correlation property of the two-mode resource state expressed in Eq. (31). This can also be used to indirectly confirm security of teleportation channel Alice and Bob possess. If the two-mode resource state they have exhibits correlation in Eq. (31) larger than NCB, i.e. $F_n>F_{\rm nc}$, their communication channel is secure. For instance, Ref. \cite{bib:PhysRevLett.108.030503} describes how to measure an arbitrary functional form of EPR correlation like Eq. (31) using homodyne detections.

For the case of using TMSV as a resource, Eq. (\ref{eq:Foperator}) can be readily caluculated to give the teleportation fidelity as 
\begin{eqnarray}
F_{\rm tele, TMSV}=\frac{\Gamma(n+\frac{1}{2})  _2F_1\left(\frac{1}{2},-n;\frac{1}{2}-n;(1-\frac{2}{1+e^{-2r}})^2\right)}{\sqrt{\pi}(1+e^{-2r})n!} 
\end{eqnarray}
 with $r$ squeezing parameter.

For other resource states, we can use density matrix elements of $ \hat{\mathcal{F}}_n $ in Fock-state basis. In Ref. \cite{bib:PhysRevA.95.052343}, we obtained the explicit expression for $ n = 0 $. The matrix element $\bra{i,j} \hat{\mathcal{F}}_0 \ket{l,m}$ is nonzero only if $j-i=m-l$, that is, the difference of number between two-mode states is the same both in bra and ket. Let us set this same difference to be $d\equiv j-i=m-l$. Then, we have
	\begin{align} \label{eq:F0element}
	\bra{i,j} \hat{\mathcal{F}}_0 \ket{l,m} & = \left( \mathcal{F}_0^{(d)} \right)_{i,l} \delta_{d, j-i} \delta_{d, m-l} , \\
	\textrm{with } \left( \mathcal{F}_0^{(d)} \right)_{i,l} & \equiv \frac{(i+l+d)!}{2^{i+l+d+1} \sqrt{i! (i+d)! l! (l+d)!}} , \nonumber
	\end{align}
where $ \delta_{i,j} $ is the Kronecker delta function. When $ n $ is nonzero, density matrix elements can be expressed in terms of $ ( \mathcal{F}_0^{(d)} )_{i,l} $. For $ n = 1 $, using $ L_1 ( \hat{O}_\textrm{EPR} ) \ket{l,m} = (l+m) \ket{l,m} - \sqrt{lm} \ket{l-1,m-1} - \sqrt{(l+1)(m+1)} \ket{l+1,m+1} $, we have
	\begin{align} \label{eq:F1element}
	\bra{i,j} \hat{\mathcal{F}}_1 \ket{l,m} = & \left( \bra{i,j} L_1 ( \hat{O}_\textrm{EPR} ) \right) e^{-\hat{O}_\textrm{EPR}} \left( L_1 ( \hat{O}_\textrm{EPR} ) \ket{l,m} \right) \nonumber \\
	= & \left( \mathcal{F}_1^{(d)} \right)_{i,l} \delta_{d, j-i} \delta_{d, m-l} , 
	\end{align}
with
 \begin{align}
	\left( \mathcal{F}_1^{(d)} \right)_{i,l} \equiv & (2i+d) \left[ (2l+d) \left( \mathcal{F}_0^{(d)} \right)_{i,l} - \sqrt{l(l+d)} \left( \mathcal{F}_0^{(d)} \right)_{i,l-1} - \sqrt{(l+1)(l+1+d)} \left( \mathcal{F}_0^{(d)} \right)_{i,l+1} \right] \nonumber \\
	& - \sqrt{i(i+d)} \left[ (2l+d) \left( \mathcal{F}_0^{(d)} \right)_{i-1,l} - \sqrt{l(l+d)} \left( \mathcal{F}_0^{(d)} \right)_{i-1,l-1} - \sqrt{(l+1)(l+1+d)} \left( \mathcal{F}_0^{(d)} \right)_{i-1,l+1} \right] \nonumber \\
	& - \sqrt{(i+1)(i+d+1)} \left[ (2l+d) \left( \mathcal{F}_0^{(d)} \right)_{i+1,l} - \sqrt{l(l+d)} \left( \mathcal{F}_0^{(d)} \right)_{i+1,l-1} - \sqrt{(l+1)(l+1+d)} \left( \mathcal{F}_0^{(d)} \right)_{i+1,l+1} \right] . \nonumber
	\end{align}
The expression for $ \bra{i,j} \hat{\mathcal{F}}_2 \ket{l,m} $ can be similarly obtained.

We here find the optimal resource state for teleportation of DFSs. The problem is to find the maximum eigenvalue of $ \hat{G}_n^\lambda \equiv \hat{\mathcal{F}}_n - \frac{\lambda}{2} ( \hat{n}_\textrm{A} + \hat{n}_\textrm{B} ) $ where $ \lambda $ is a Lagrange multiplier and $ \hat{n}_{A(B)} $ is the photon number operator of mode $ A(B) $. For fixed $ \lambda $, the eigenstate with the largest eigenvalue, say $ \ket{\Psi_n^{\lambda,\textrm{max}}} $, becomes the optimal state which leads to the highest fidelity among the states with the same photon number $ n_\textrm{av} = \frac{1}{2} \langle \hat{n}_\textrm{A} + \hat{n}_\textrm{B} \rangle $ \cite{bib:PhysRevA.95.052343}. We numerically investigate the eigenvalue of $ \hat{G}_n^\lambda $ by varying $ \lambda $ in the truncated Fock basis.

Interestingly, in Eqs. (\ref{eq:F0element}) and (\ref{eq:F1element}), we find that the fidelity operator $ \hat{\mathcal{F}}_n $ is in a block-diagonal form, where each block corresponds to the space spanned by $ \ket{i,i+d} $ with the same photon number difference $ d $. Eigenstates reside in each block subspace and thus we can find the optimal states in a subspace with a fixed $ d $. For $ d = 0 $, eigenstates become photon-number-entangled states (PNESs) of the form $ \sum_i c_i \ket{i,i} $, which is shown to be optimal below. States with nonzero $ d $ have the average photon number $ n_\textrm{av} \ge \frac{|d|}{2} $, which makes them inefficient especially in the small-photon regime. In Fig. \ref{fig:dfidelity}, we show the optimal teleportation fidelity achieved with the states with $ d = 0, 1, 2 $.
	\begin{figure}[t]
	\centering \includegraphics[clip=true, width=0.8\columnwidth]{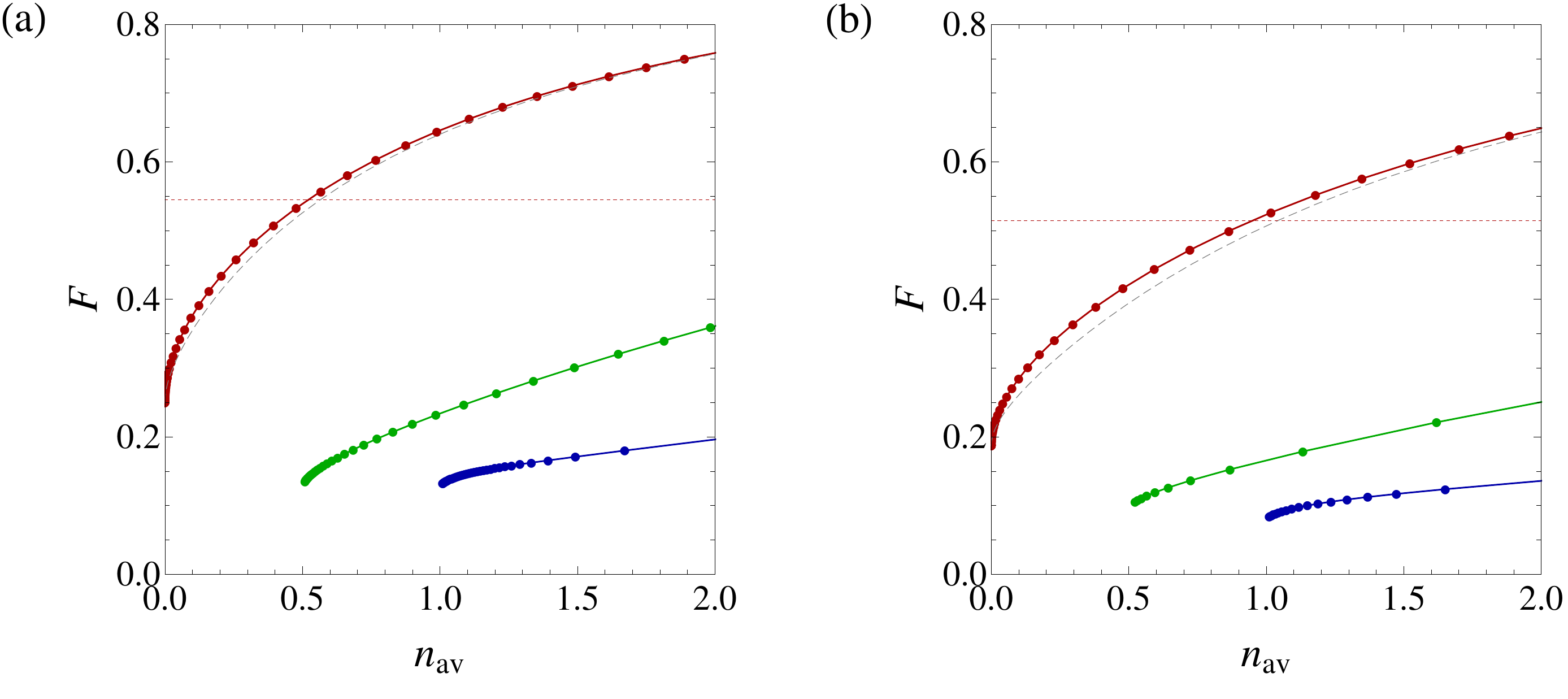}
	\caption{\label{fig:dfidelity} Teleportation fidelity as a function of average photon number using non-Gaussian resource states with $ d = $ 0 (red), 1 (green), and 2 (blue) from top to bottom, for the case of input states with $ n = $ (a) 1 and (b) 2. The gray dashed curve represents teleportation fidelity with TMSV. The horizontal dotted line represents the genuine no-cloning bound.}
	\end{figure}
It is shown states with $ d = 1, 2 $ achieve much smaller fidelity than that with PNES and they cannot even reach the no-cloning bound with small photon number $ n_\textrm{av} \le 2 $. For $ d < 0 $, we can obtain the same result as the case of positive $ d $ with the same magnitude $ |d| $ by changing roles of two modes, because the operator $ \hat{\mathcal{F}}_n $ is symmetric under permutation of each mode. We also plot the fidelity achieved with TMSV in Fig. \ref{fig:dfidelity}, and it is observed that the optimal PNES slightly outperforms TMSV.

As we have now obtained the optimal resources for each case in CV teleportation, we can make a more rigorous statement concerning the condition to achieve secure quantum teleportation. In Fig. \ref{fig:NGfidelity}, we plot the teleportation fidelity of the optimal PNES compared with that of TMSV and the required photon number to achieve the no-cloning bound.
	\begin{figure}[t]
	\centering \includegraphics[clip=true, width=0.4\columnwidth]{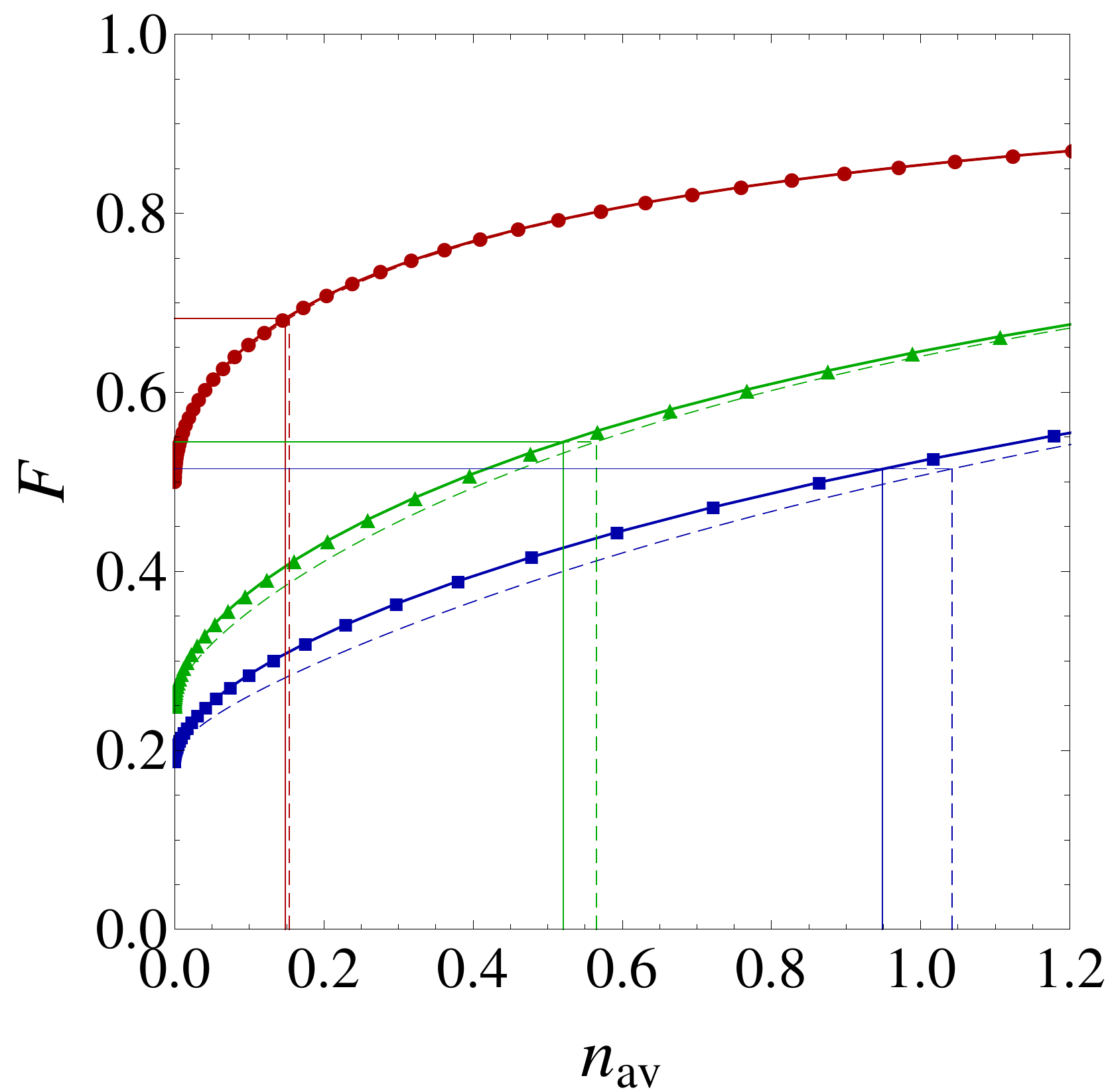}
	\caption{\label{fig:NGfidelity} Teleportation fidelity with PNES. Symbols represent the fidelity for $ n = $ 0 (red circles), 1 (green triangles), 2 (blue squares). Dashed curves represent the fidelity with TMSV for $ n = $ 0 (red), 1 (green), 2 (blue) from top to bottom. Horizontal lines represent the genuine no-cloning bound for each input state. The vertical solid lines show the required photon number of PNES and the vertical dashed lines show that of TMSV.}
	\end{figure}
For the coherent-state input, the fidelities with Gaussian resource and non-Gaussian resource are very close, that is, non-Gaussian improvement is negligible. On the other hand, the non-Gaussian improvement is comparably large for the DFS input. The required photon number for secure teleportation increases with $n$ as $ n_\textrm{av} \simeq $ 0.148, 0.521, and 0.943, respectively for $ n = 0, 1, 2 $. While the teleportation fidelity is improved by employing non-Gaussian resources, we still draw the same conclusion as in main text that quantum non-Gaussian input states require more resources to achieve secure quantum teleportation.

\section*{S7. Relevance of overlap measure for mixed-state inputs}
In main text, we have used the overlap measure ${\rm Tr} \{\rho\tau\}$  for mixed-state inputs, as an extension of the usual fidelity measure for pure-state inputs, to quantify the {\it closeness} of two states. We have also raised an undesirable feature regarding this measure, namely, given $\rho=0.75|0\rangle\langle0|+0.25|1\rangle\langle1|$, a different state $\tau=0.9|0\rangle\langle0|+0.1|1\rangle\langle1|$ may yield a higher overlap than $\rho$ itself, i.e. ${\rm Tr} \{\rho\tau\}>{\rm Tr} \{\rho^2\}$. This happens because the state $\tau$ has a higher purity than $\rho$, i.e. ${\rm Tr} \{\tau^2\}>{\rm Tr} \{\rho^2\}$. A possible remedy for this problem is to take a proper normalization of measure, i.e.  $M\equiv\frac{{\rm Tr} \{\rho\tau\}}{\sqrt{{\rm Tr} \{\rho^2\}{\rm Tr} \{\tau^2\}}}$. Obviously, $0\le M\le1$ takes a maximum value 1 if and only if $\rho=\tau$. However, this normalized measure is nonlinear with respect to states, thus it becomes very demanding to obtain a numerically stable cloning bound.

On the other hand, our overlap measure is still relevant to capture the essence of quantum cloning and related problems. 
While the undesirable higher values of overlap may indeed occur at single instances, we note that quantum cloning deals with a set of {\it infinitely many different} input states, and importantly, their {\it average fidelity} (overlap). Let us look into this issue based on the phase-space overlap ${\rm Tr} \{\rho\tau\}=\pi\int d^2\alpha W_\rho(\alpha)W_\tau(\alpha)$.  In general, we may point out three factors determining this overlap---center, shape, and purity of each profile. (Of course, shape and purity are not completely independent of each other.) There is a kind of trade-off relation among these factors: While one attempts to score higher by making an output with a higher purity, one should also make sure that the output profile has the same center (complex amplitude) as the target state. If the centers do not coincide, the overlap becomes significantly reduced. It means, for instance, that a random preparation of pure states is not helpful and that one should systematically gain information on the unknown amplitude distributed over the entire phase space. The latter efforts is the very essence in our study of cloning. We explore the minimum possible noise that must be intrinsically added during the process of information gain, whether quantum or classical cloning, which necessarily reduces the purity of the output state.

We have proved in Methods of main text that a covariant cloner is the optimal machine if the trace overlap is taken as a figure of merit. We here point out that the proof on the optimality of covariant cloner can be readily extended to other measures too, e.g. Hilbert-Schmidt distance ${\rm Tr}\{(\rho-\tau)^2\}$. This covariant cloner, and actually any conceivable cloners, must introduce an intrinsic amount of added noise, which yields a no-cloning bound in each measure.

	\begin{figure}[t]
	\centering \includegraphics[clip=true, width=0.4\columnwidth]{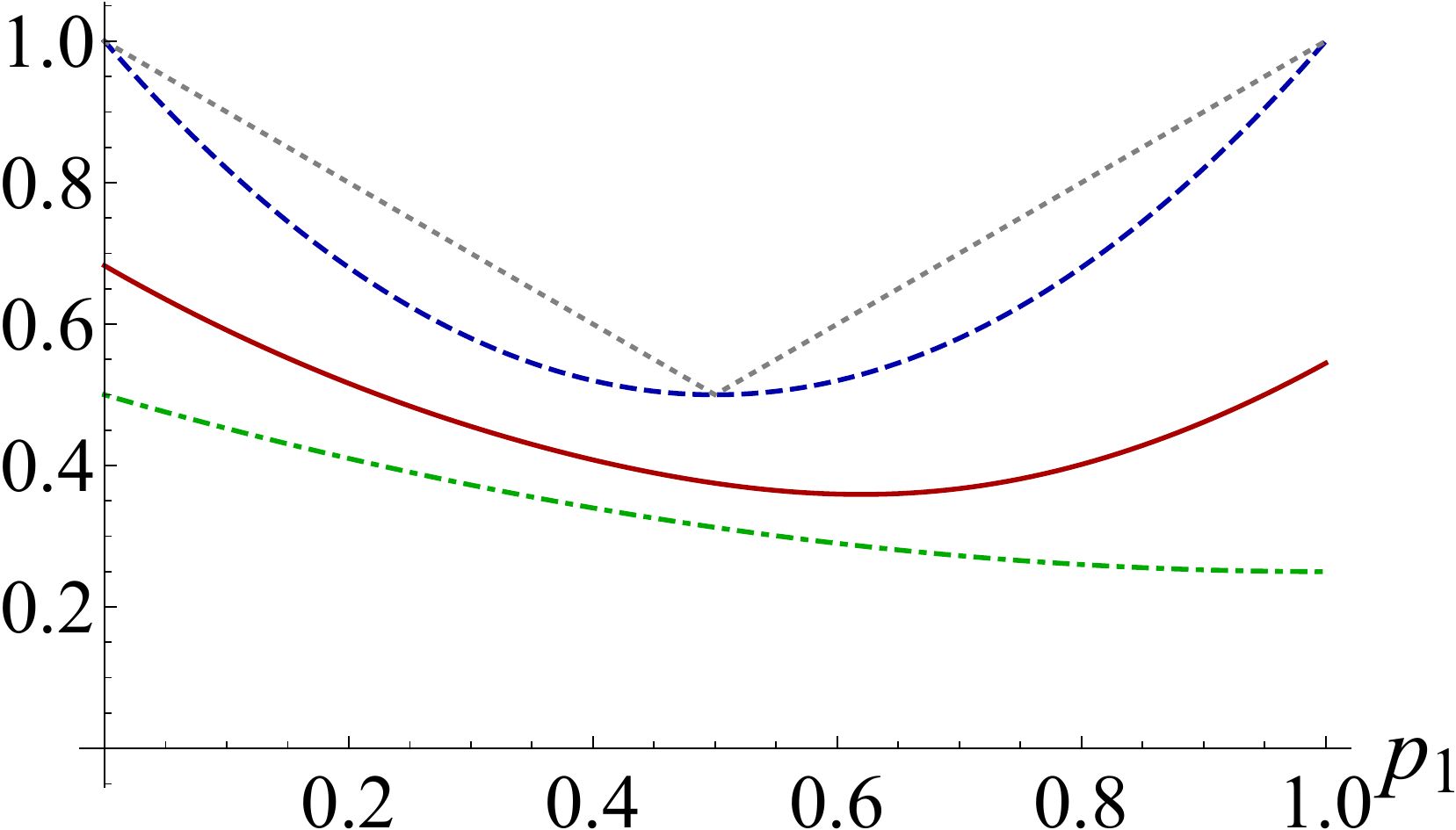}
	\caption{\label{fig:hypo} Values of ${\rm Tr} \{\rho\tau\}$ as a fuction of $p_1$ for an input $\rho=(1-p_1)|0\rangle\langle0|+p_1|1\rangle\langle1|$. Gray: a hypothetical value maximized over $\tau$ for each input $\rho$, i.e. ${\rm Max}_\tau {\rm Tr} \{\rho\tau\}={\rm Max}\{p_1,1-p_1\}$, Blue: ${\rm Tr} \{\rho^2\}$, Red curve (no-cloning bound) and green curve (classical bound) represent the average fidelity over all displacements in phase space for each input $\rho$. }
	\end{figure}

In Fig. \ref{fig:hypo}, we show the case of the mixed state $\rho=(1-p_1)|0\rangle\langle0|+p_1|1\rangle\langle1|$ with relevant quantities including quantum no-cloning bound (red) and classical bound (green). 
Our task is in a sense to estimate the unknown displacement $\alpha$ of the input state $D(\alpha)\rho D^\dag(\alpha)$ for which $\alpha$ is randomly distributed over the entire phase space. This is why high values such as the one by an optimal, hypothetical, pure state (grey dotted) and even the one by the input-state itself (blue dashed) cannot be obtained in the ensemble sense of average overlap. For instance, the grey dotted line would be achieved when the output state is $D(\alpha)|0\rangle$ or $D(\alpha)|1\rangle$ for the input $D(\alpha)\rho D^\dag(\alpha)$, i.e., the same displacement obtained systematically for each case, which is an impossible task. 
Therefore, if one demonstrates that the value of overlap is larger than NCB after taking average over many different input states, one can safely say that the quality of similarity between input and output states is legitimately good ruling out the possibility that the high score originates from undesirable feature of the measure.

\end{widetext}

\end{document}